\newcommand{\msun}{\hbox{${M}_{\odot}$}}

\newcommand{\simgt}{\lower 2pt \hbox{$\, \buildrel {\scriptstyle >}\over {\scriptstyle\sim}\,$}}
\newcommand{\simlt}{\lower 2pt \hbox{$\, \buildrel {\scriptstyle <}\over {\scriptstyle\sim}\,$}}

\documentclass{emulateapj}
\usepackage{natbib,psfig,amsmath}

\begin{document}
\shorttitle{Deep GALEX Observations of the Coma Cluster: Source Catalog and Galaxy Counts}
\shortauthors{Hammer et al.}
\submitted{Accepted by the Astrophysical Journal Supplement Series 04/08/2010}
\title{Deep GALEX Observations of the Coma Cluster: Source Catalog and Galaxy Counts}
\author{D.~Hammer$^{1,2}$,~A.~E.~Hornschemeier$^{2,1}$,~B.~Mobasher$^{3}$,~N.~Miller$^{4,1}$,~R. Smith$^{5}$,~S.~Arnouts$^{6}$,~B.~Milliard$^{7}$, and ~L.~Jenkins$^{2,1}$}
\altaffiltext{1}{Department of Physics and Astronomy, The Johns Hopkins University, 3400 N. Charles Street, Baltimore, MD 21218, USA}
\altaffiltext{2}{Laboratory for X-ray Astrophysics, Code 662, NASA GSFC, Greenbelt, MD 20771, USA}
\altaffiltext{3}{Department of Physics and Astronomy, University of California, Riverside, CA 92521, USA}
\altaffiltext{4}{Department of Astronomy, University of Maryland, College Park, MD, 20742-2421, USA}
\altaffiltext{5}{Department of Physics, University of Durham, South Road, Durham DH1 3LE, UK}
\altaffiltext{6}{Canada-France-Hawaii Telescope Corporation, 65-1238 Mamalahoa Hwy, Kamuela, Hawaii 96743, USA}
\altaffiltext{7}{Laboratoire d'Astrophysique de Marseille, BP 8, Traverse du Siphon, 13376 Marseille Cedex 12, France}

\begin{abstract}
We present a source catalog from deep 26 ks {\it GALEX} observations of the Coma cluster in the far-UV (FUV; 1530 \AA) and near-UV (NUV; 2310 \AA) wavebands.
The observed field is centered $\sim$0.9$^{\circ}$ (1.6 Mpc) south-west of the Coma core,
and has full optical photometric coverage by SDSS and spectroscopic coverage to $r$$\sim$21.
The catalog consists of 9700 galaxies with GALEX and SDSS photometry,
including 242 spectroscopically-confirmed Coma member galaxies that range from giant spirals and elliptical galaxies to dwarf irregular and early-type galaxies.
The full multi-wavelength catalog (cluster plus background galaxies) is $\sim$80\% complete to NUV=23 and FUV=23.5, and has 
a limiting depth at NUV=24.5 and FUV=25.0 which corresponds to a star formation rate of 10$^{-3}$ $\msun$ yr$^{-1}$ at the distance of Coma.
The {\it GALEX} images presented here are very deep and include detections of many resolved cluster members superposed on a dense field of unresolved background galaxies.
This required a two-fold approach to generating a source catalog: we used a Bayesian deblending algorithm to measure faint and compact sources (using SDSS coordinates as a position prior),
and used the {\it GALEX} pipeline catalog for bright and/or extended objects.
We performed simulations to assess the importance of systematic effects (e.g.~object blends, source confusion, Eddington Bias) that influence source detection and photometry when using both methods.
The Bayesian deblending method roughly doubles the number of source detections and provides reliable photometry to a few magnitudes deeper than the {\it GALEX} pipeline catalog.
This method is also free from source confusion over the UV magnitude range studied here; conversely, we estimate that the
{\it GALEX} pipeline catalogs are confusion limited at $NUV$$\sim$23 and $FUV$$\sim$24.
We have measured the total UV galaxy counts using our catalog and report a $\sim$50\% excess of counts across $FUV$=22-23.5 and $NUV$=21.5-23 relative to previous {\it GALEX} measurements,
which is not attributed to cluster member galaxies.
Our galaxy counts are a better match to deeper UV counts measured with HST.
\end{abstract}
\keywords{galaxies: clusters: individual (Coma)  - galaxies: statistics - catalogs - techniques: photometric - ultraviolet: galaxies}

\pagebreak
\section{Introduction}
The broadband ultraviolet (UV; 1000-3000 \AA) emission from galaxies is a unique age-dating tool because it traces stellar populations at both very early and very
late phases of evolution for different galaxy types.
The far-UV (FUV; 1000-2000 \AA) and near-UV (NUV; 2000-3000 \AA) emission from normal star-forming galaxies is typically produced by short-lived intermediate-mass stars on the main sequence
(t$\simlt$1 Gyr; 2$\simlt$$M_{ZAMS}$$\simlt$5 \msun), which provides a direct measure of the star formation rate \citep{KennicuttSF}.
In contrast, the FUV emission from {\it quiescent} galaxies is associated with low-mass He-burning stars that evolve through the horizontal branch
\citep[$M_{ZAHB}$$\sim$0.5 $\msun$;][]{Greggio1990,Dorman1995,Park1997}; these stars produce an upturn in the galaxy spectrum shortward of 2000 \AA~
\citep[e.g.~UV-excess (UVX) or UV-upturn galaxies;][]{Code1969,Brown1995,Oconnell1999}.
The NUV emission from quiescent galaxies is typically dominated by the hottest main-sequence turnoff stars and subgiants \citep[e.g.][]{Dorman2003}.

The Coma cluster, Abell 1656, is one of the nearest ($z$$\sim$0.023) examples of a rich galaxy cluster
\citep[Abell Class 2 - Bautz \& Morgan Type II cluster;][]{Abell1989,Bautz1970} and is among the best studied
local galaxy clusters due to its accessibility at high Galactic latitude ($b$$\sim$88$^{\circ}$).
UV observations of the Coma cluster have been relatively rare as compared to optical wavelengths owing to the need for space-based facilities to avoid atmospheric absorption, 
the inability of early UV instruments to obtain both deep and wide-field coverage of nearby clusters, and unfortunate instrument failures during previous observing campaigns
(e.g.~FAUST during the Astro-1 shuttle mission, and the UIT NUV camera during Astro-2).
The balloon-borne FOCA instrument (2000 \AA) performed the best previous UV imaging surveys with sufficient sensitivity, angular resolution (20$\arcsec$), and field of view
(circular FOV of 2$^{\circ}$ diameter) to detect a statistically significant sample of galaxies in local clusters \citep{Milliard1991}.
FOCA performed a 3 ks observation at the center of the Coma cluster \citep{Donas1991} that detected optically-bright early-type galaxies in the cluster ($M_{B_{j}}$$\simlt$-19),
and fainter star forming galaxies \citep[$M_{B_{j}}$$\simlt$-16; SFR$\simgt$ 0.05 \msun~yr$^{-1}$ following][]{KennicuttSF}.
This observation resulted in the first ever UV luminosity function (LF) measured for a galaxy cluster \citep{Andreon1999, Cortese2003a,Cortese2003b}.


Launched in 2003, the {\it Galaxy Evolution Explorer} \citep[{\it GALEX};][]{Martin2005} significantly improved the capability for performing
UV surveys due to its relatively high sensitivity, good spatial resolution ($\sim$5$\arcsec$ FWHM), and large circular FOV (1.25$^{\circ}$ diameter).
{\it GALEX} has observed several local galaxy clusters thus far, from both its Guest Investigator program \citep[e.g.~NFPS clusters;][]{Rawle2008} and also via its imaging survey mode.
For example, {\it GALEX} All-Sky Imaging Survey observations with 0.1 ks exposures detected a large sample of dwarf early-type galaxies in the nearby (20 Mpc) Virgo cluster \citep[][]{Boselli2005b}.
{\it GALEX} has also performed moderate-depth (1.3-3.5 ks) observations of the Abell 1367 and Coma clusters both located at $\sim$100 Mpc.
Specifically, {\it GALEX} observed a single field just north of center for A1367, while Coma has extensive wide-field coverage that consists of 11 {\it GALEX} fields that
span from the periphery of the core to the cluster virial radius and cover 27 Mpc$^{2}$.
Both sets of observations allowed for the detection of dwarf star forming galaxies (SFR$\simgt$0.02 \msun yr$^{-1}$) and intermediate-mass early-type galaxies (M$_{*}$$>$10$^{10}$ \msun), 
and resulted in the deepest UV LFs that have been measured for local galaxy clusters thus far \citep[M$_{UV}$$<$-14;][]{Cortese2005,Cortese2008}.

We have obtained a deep 30 ks {\it GALEX} observation of an off-center region of the Coma cluster (`Coma-3'), which compliments the existing wide-field coverage with {\it GALEX}.
These data are 95\% complete to m$_{UV}$$\sim$24.5 in the {\it GALEX} FUV and NUV filters ($\sim$3 magnitudes deeper than the wide-field {\it GALEX} coverage),
which corresponds to a star formation rate of SFR$\sim$10$^{-3}$ \msun~yr$^{-1}$ for galaxies at the distance of Coma.
The depth of our observation also allows for the detection of quiescent dwarf early-type (dE) galaxies that were too faint to be detected in previous UV surveys of the Coma cluster.

The `Coma-3' field \citep[][]{Komiyama2002} is located $\sim$1$^{\circ}$ (1.75 Mpc) south-west of the cluster center.
This location has been the subject of several studies since the discovery of a secondary peak in the cluster X-ray emission located at the northern edge of the Coma-3 field \citep{White1993}.
The X-ray emission is associated with the infalling NGC 4839 galaxy group \citep[e.g.][]{Colless1996,Neumann2001,Briel2001}, and has been linked 
to a population of young/post-starburst galaxies that may trace the X-ray substructure \citep[e.g.][]{Caldwell1993,Caldwell1998,Poggianti2004,Smith2009}.
This field has extensive panchromatic coverage from X-ray to radio wavelengths \citep[a complete review is presented in][]{Miller2009}, including broadband optical and spectroscopic coverage
by the {\it Sloan Digital Sky Survey} \citep[SDSS;][]{Adelman2008}, and deeper optical spectroscopic observations with the William Herschel Telescope and MMT Hectospec that 
provide redshift coverage to $r$$\sim$21 \citep[R.~Marzke et al.~2010, {\it in prep};][]{Mobasher2001}.


This paper describes our deep {\it GALEX} observation in the Coma cluster with full optical coverage by the SDSS (\S2).
A galaxy catalog is constructed by taking UV measurements from both the {\it GALEX} pipeline catalog and also using a Bayesian deblending technique to measure the UV-faint population,
and we use simulations to determine the detection and photometry limits for both methods (\S3).
The catalog is used to measure the UV galaxy counts and to estimate the {\it GALEX} source confusion limits (\S4).
The source catalog will be used in a companion paper to measure a deep UV LF for the Coma cluster and to study the stellar populations of Coma members.
We assume Coma is located at a distance of 100 Mpc, which corresponds to a distance modulus of DM$=$35.0 ($z=0.023$) for H$_{0}$$=71$~km s$^{-1}$Mpc$^{-1}$,
$\Omega_{\lambda}$=0.73, and $\Omega_{m}$=0.27.


\section{Data}
\subsection{GALEX Data}
{\it GALEX} images the sky simultaneously in two broadband filters centered at 1539\AA~(FUV) and 2316\AA~(NUV).
The {\it GALEX} pipeline produces images with a circular field of view of 1.25$^{\circ}$ diameter. 
However, we only consider sources located within the inner 0.6$^{\circ}$ {\it GALEX} FOV to avoid
artifacts (bright star glints) that affect the NUV band at locations further from the image center.
The {\it GALEX} plate scale is 1.5$\arcsec$ pixel$^{-1}$ which adequately samples the $\sim$4.5-7$\arcsec$ PSF (FWHM).
We refer the reader to \cite{Martin2005} and \cite{Morrissey2005, Morrissey2007} for a more detailed description of the {\it GALEX} instrumentation and calibration.

We were awarded a deep 30 ks {\it GALEX} observation of the Coma cluster as part of the Cycle-II Guest Investigator program (P.I.~Hornschemeier).
This field has an integrated exposure time similar to the Deep Imaging Survey (DIS), which are some of the deepest observations taken by the {\it GALEX} mission.
The observing sequence for our {\it GALEX} field consists of 35 individual exposures with durations that range between 100-1700 seconds.
 The analysis presented in this study was performed on $\sim$26 ks images in both {\it GALEX} bands, which were the deepest images available when this study was initiated.
The {\it GALEX} field is centered at R.A.=194$^{\circ}$.34167 and Dec.=27$^{\circ}$.17722 (J2000),
which is located 0.9$^{\circ}$ (1.6 Mpc) south-west of the Coma center in a region of the cluster known as `Coma-3' \citep{Komiyama2002}.
The entire {\it GALEX} field is located inside the virial radius of the Coma cluster as shown in Figure 1 \citep[r$_{vir}$=2.9 Mpc;][]{Lokas2003}.

Object detection and photometry is not trivial for deep {\it GALEX} images.
The {\it GALEX} pipeline provides source catalogs using the SExtractor software \citep{Bertin96}.
However, the high source density in deep images and the relatively large {\it GALEX} PSF result in a non-negligible fraction of
object blends in the pipeline catalogs, i.e.~two or more objects in close proximity on the sky are considered a single source.
The {\it GALEX} pipeline catalogs are also limited at faint magnitudes by source confusion.
\cite{Xu2005} estimated that source confusion leads to systematic detection and photometric biases for objects fainter than $NUV$$\approx$24 and $FUV$$\approx$25.3.
In fact, we will argue in \S4.2 that the {\it GALEX} pipeline catalogs are confusion limited at brighter magnitudes than reported in \cite{Xu2005}.



\cite{Morrissey2007} suggest that alternative methods of object detection and photometry are required to
separate sources in deep {\it GALEX} images and to avoid the effects of source confusion.
For instance, \cite{Zamojski2007} applied a Bayesian deblending method to {\it GALEX} DIS images of the COSMOS field (z$\sim$0.7).
The authors measured the UV flux by fitting a PSF profile to sources that were identified in high-resolution optical images (the optical coordinates are used as a position prior).
Although a point source approximation is reasonable for the faint {\it field} galaxy population observed by {\it GALEX}, it does not describe the foreground cluster
members that are resolved (or at least marginally resolved) in the Coma field studied here.
We adopt a hybrid of these methods, using SExtractor to measure bright extended sources and Bayesian deblending
for the unresolved population (using SDSS coordinates as a position prior).
A description of these two methods and a demonstration of their reliable limits are presented in \S3.

We measured a systematic offset in the {\it GALEX} image coordinates relative to SDSS after comparing the positions of 200 stars detected in both imaging datasets.
The {\it GALEX} coordinates are offset to the north-east of SDSS (on average) by 0.4$\arcsec$ and 0.6$\arcsec$ in R.A. and Dec., respectively.
We use the corrected {\it GALEX} coordinates (i.e.~shifted to match the SDSS system) in this study.
The {\it GALEX} magnitudes presented hereafter have been corrected for foreground Galactic extinction
using the \cite{Schlegel1998} reddening maps and the ratio of extinction to reddening reported in \citet[][$A$$_{fuv}$/E(B$-$V)=8.24, $A$$_{nuv}$/E(B$-$V)$\simeq$8.20]{Wyder2007}.
Coma is located at high Galactic latitude ($b\sim88^{\circ}$), resulting in typical Galactic extinction values of only $A$$\sim$0.08 mag in both {\it GALEX} filters.

\subsection{Optical Data}
The Sloan Digital Sky Survey \citep[SDSS;][]{York2000} has observed the entire Coma cluster in the broadband {\it ugriz} filters with a median seeing
condition of 1.4$\arcsec$ in the $r$-band. The data we use are taken from the SDSS DR6\footnote[6]{SDSS photometry for the Coma cluster have
been reprocessed as part of DR7. A comparison of DR6 and DR7 sources
in the same Coma field studied here reveal only a negligible difference in their magnitudes and source positions.}
which is 95\% complete to $r$=22.2 \citep{Adelman2008}.
We use {\tt PETROMAG} photometry to estimate the total flux of galaxies, and {\tt MODELMAG} measurements for optical color analysis.
Magnitudes were corrected for Galactic reddening using the {\tt extinction} values available in the SDSS PhotoObj tables.
We also applied a small magnitude correction to match the AB system (-0.036, 0.012, 0.010, 0.028, and 0.040 for the {\it ugriz} filters, respectively),
and imposed a minimum photometry error for all SDSS magnitudes \citep[$\Delta$m=0.05,0.02,0.02,0.02,0.03 for the $ugriz$ filters, respectively;][]{Blanton2007}.
We do not include SDSS sources fainter than $r$=24 as we found $\sim$40\% of these objects are spurious detections.  This was determined by comparing SDSS detections to 
optical data with much higher spatial resolution and greater depth from HST-ACS imaging of the Coma cluster \citep[see Appendix A for details;][]{Carter2008}.


We rely on the SDSS DR6 to identify objects in the {\it GALEX} pipeline catalog and for producing accurate position priors for the Bayesian deblending analysis.
Therefore it is essential to limit the number of false sources in the SDSS catalog.
False sources in the SDSS catalog typically result from `shredding' of bright galaxies, i.e.~a single galaxy is detected as two or more separate objects \citep{Abaz2004}.
In order to identify false sources, we have visually inspected SDSS detections brighter than $r$=22.5 that are matched to the {\it GALEX} pipeline catalog ($\sim$4000 galaxies);
there are $\sim$27000 SDSS sources in our field thus we limited our inspection to galaxies that are likely to have a UV counterpart.
We removed 181 sources from the DR6 catalog that have characteristics typical of shredding
(e.g.~no clear separation from the parent galaxy, unphysical colors, magnitudes that are significantly brighter than expected based on visual inspection),
and added the flux to the parent galaxy in the SDSS DR6 catalog. Note that a relatively high fraction of shredded sources are associated with Coma member galaxies,
particularly for low surface brightness (LSB) galaxies and also within the halos of bright early-type galaxies.
In Figure \ref{sdss_blends} we provide typical examples of Coma member galaxies that were shredded into several sources.
We have also removed 19 SDSS objects that were classified as galaxies but we identified as diffraction spikes due to bright stars.

We must rely on the star/galaxy (S/G) classification from SDSS DR6 (i.e.~the `type' parameter) to identify galaxies as {\it GALEX} lacks the spatial resolution to perform this task.
The HST-ACS data in the Coma cluster allows us to test the accuracy of the SDSS S/G classification (Appendix B).
We found that objects classified as galaxies by SDSS are $\simgt$90\% accurate at magnitudes brighter than $r$=24.0.
However, over one-half of all objects classified as stars by SDSS with $r$$>$21 are actually galaxies.
Therefore, we use a SDSS catalog that consists of objects with $r$$<$21 that are identified as galaxies in SDSS DR6,
and all objects fainter than $r$=21 (regardless of their SDSS S/G classification) to avoid excluding many real galaxies at faint magnitudes.
The final SDSS catalog consists of 26558 (18431) objects with $r$$\leq$24.0 that are located within 0.6$^{\circ}$ (0.5$^{\circ}$) from the center of the {\it GALEX} field.

\section{Properties of the UV Catalogs}
In this section we describe the UV source catalogs used in this study, and use simulations to establish their source detection and photometry limits.
We use two separate methods of UV source detection and photometry.
We take UV flux measurements from the {\it GALEX} pipeline catalogs for bright and/or extended sources that are located within the inner 0.6$^{\circ}$ {\it GALEX} FOV.
We use a Bayesian deblending method \citep[e.g.][]{Guillaume2006} to measure faint, unresolved objects located within 0.5$^{\circ}$ from the center of the {\it GALEX} field.

\subsection{Bayesian Deblending Catalog}
In their {\it GALEX} calibration paper, \cite{Morrissey2007} suggested that source detection and photometry for deep {\it GALEX} images
may be significantly improved by PSF-fitting the UV flux for objects detected in higher-resolution optical imaging.
We have applied this technique (referred to as `Bayesian deblending' throughout this paper) to our deep {\it GALEX} image.
Performing source detection in the high-resolution optical image allows us to:
(a) separate objects in close proximity on the sky that would otherwise be blended into a single source, and
(b) avoid systematic effects that influence source detection and photometry at faint magnitudes \citep[e.g.~source confusion;][]{Hogg2001}.
This method is somewhat akin to performing SExtractor dual-image color photometry (i.e.~UV flux is measured inside apertures defined in the optical image),
however, SExtractor was not designed to separate objects in crowded images.
Bayesian deblending allows for crowded-field photometry by exploiting the known location of objects from higher resolution optical images, the measured {\it GALEX} PSF,
and the fact that most sources are unresolved in {\it GALEX} images.

We have created a UV source catalog using custom software (developed by {\it GALEX} team members and led by co-authors S.A.~and B.M.)
that is based on the Bayesian deblending technique described in \citet[][]{Guillaume2006},
which was designed specifically to measure faint compact objects in crowded {\it GALEX} images.
This software has been used in several studies with deep {\it GALEX} imaging \citep[e.g.][]{Zamojski2007,Salim2009,Ibert2009}, and
a brief description and demonstration is also provided at the {\it GALEX} GI website\footnote[7]{www.galex.caltech.edu/researcher/techdoc-ch5a.html}.
To summarize, the method constructs a parametric model for the UV image that consists of both a source flux and background component.
The source flux is modeled by taking the optical coordinates and flux values from our SDSS catalog as a starting reference (i.e.~a prior), and 
then assumes the corresponding UV light distribution follows the {\it GALEX} PSF.
The background is measured following the same technique used for the {\it GALEX} pipeline \citep{Morrissey2007}.
The UV flux for each object is fit by solving the likelihood of the parametric equation using an expectation-maximisation (EM) technique.
Photometry is reliable for galaxies with apparent sizes smaller than $\sim$10$\arcsec$ (i.e.~$\sim$2$\times$FWHM).
For more extended galaxies, the flux will be underestimated due to a UV light profile that is much broader than the PSF.
We therefore rely on the {\it GALEX} pipeline catalog (\S3.2) for extended galaxies.

Bayesian deblending requires an accurate PSF model for both {\it GALEX} filters.
As such, we have restricted the our analysis to the inner 0.5$^{\circ}$ {\it GALEX} FOV to avoid regions
where the PSF is difficult to model owing to its asymmetry \citep{Morrissey2007}.
This is a conservative cutoff since the {\it GALEX} PSF tends to be more symmetric for images
that are constructed by stacking observations taken at many roll angles (which is the case for our {\it GALEX} image).
We have measured the PSF for both {\it GALEX} filters by stacking isolated stars that are brighter than $NUV$=$FUV$=18 and located within 0.5$^{\circ}$ of the image center.
We measured a 4.8$\arcsec$ and 5.3$\arcsec$ FWHM for the FUV and NUV filters, respectively. The encircled energy curves of the {\it GALEX} PSFs are shown in Figure \ref{galex_psf}.
The full width at half-energy is 5.5$\arcsec$ for the FUV filter and 6.0$\arcsec$ for the NUV filter, which are similar to the average PSFs measured for fields in the {\it GALEX}
Medium Imaging Survey (MIS)\footnote[8]{Available from the {\it GALEX} Guest Investigator website http://www.galex.caltech.edu/researcher/techdoc-ch5.html\#2.}.

We have modeled the UV flux for all objects in our SDSS catalog that are located within the inner 0.5$^{\circ}$ {\it GALEX} FOV.
UV fluxes were measured for 92\% of the SDSS objects (16874/18431), of which approximately one-half
are brighter than our chosen magnitude limits of $NUV$=24.5 and $FUV$=25.0 (magnitude limits are described in \S3.3-3.4).
We visually confirmed that the majority of the unmeasured SDSS objects are too UV-faint to be detected in our {\it GALEX} images.
The EM algorithm, however, failed to converge on a flux solution for 381 objects with identifiable UV emission
for which we have taken the measurements from the {\it GALEX} pipeline catalog; the failure occurred for several reasons,
such as close proximity to a bright object or local gradients in the background level.
Note that the Bayesian deblending method is prone to assigning non-zero flux to optical priors that lack UV emission, particularly for priors located 
within the light distribution of bright extended galaxies or saturated stars.
We have estimated the fraction of such `spurious' detections by adding to the SDSS object list $\sim$4000 false priors that were 
randomly distributed across the inner 0.5$^{\circ}$ {\it GALEX} FOV.
Only $\sim$1\% (2\%) of the false priors were assigned fluxes that are brighter than our magnitude limit of $FUV$=25.0 ($NUV$=24.5).
Spurious UV detections are thus only a negligible fraction of the sources in the Bayesian deblending catalog.

We have also tested the sensitivity of the Bayesian deblending analysis to slight changes in the PSF profile, and
also for small offsets between the {\it GALEX} and SDSS coordinates,
i.e.~we re-measured sources using the PSF given for the {\it GALEX} MIS survey, and
also after shifting the {\it GALEX} and SDSS coordinates by $0.5$$\arcsec$ in each coordinate.
We conclude that the Bayesian deblending algorithm is not critically sensitive to these parameters after finding
only a negligible change to the number of source detections and their flux measurements.

\subsection{{\it GALEX} Pipeline Catalog}
The {\it GALEX} pipeline generates source catalogs using a slightly modified version of the SExtractor software, which
is adjusted to provide a more accurate estimate of the UV background \citep[][]{Morrissey2007}.
The advantages of the {\it GALEX} pipeline catalog, as compared to the Bayesian deblending catalog, are that
(a) it provides coverage across a larger area of the {\it GALEX} image (the inner 0.6$^{\circ}$ {\it GALEX} FOV versus the inner 0.5$^{\circ}$), and
(b) its photometry is relatively insensitive to the apparent sizes of galaxies, whereas the Bayesian deblending catalog underestimates the flux for extended galaxies.
We use the term `extended' throughout this paper to refer to galaxies with an optical 90\% light diameter larger than 10$\arcsec$ in the SDSS $r$-band.
The primary drawback of the {\it GALEX} pipeline catalog is that it is relatively incomplete at magnitudes fainter than $NUV$($FUV$)=21.0 (described in \S3.3).

As such, we rely on the {\it GALEX} pipeline catalog for galaxies that are brighter than $NUV$($FUV$)=21.0 and/or are extended galaxies.
We use the SExtractor {\tt MAG\_AUTO} magnitudes which are measured using elliptical Kron apertures \citep{Kron1980} and reported in the AB system.
FUV and NUV Kron magnitudes were taken from the `merged' {\it GALEX} pipeline catalog in which source detection and aperture definition are determined separately for each filter.
Note that we have visually inspected sources that meet our extended size criteria but were flagged in the SDSS DR6 as having unreliable measurements of the Petrosian light diameter ({\tt NOPETRO});
we identified 125 objects that are obviously smaller than 10$\arcsec$ for which we have taken measurements from the Bayesian deblending catalog.

Our bright/extended sample was assembled by matching the {\it GALEX} pipeline and SDSS catalogs using a standard 4$\arcsec$ matching radius \citep{Budavari2009},
and selecting 479 galaxies that match our criteria. We have added another eight galaxies that were missed by the pipeline due to object blends.
Coma member galaxies are the majority population (57\%) for the subset of these galaxies with spectroscopic redshifts (365/487).
We also include photometry from the pipeline catalog for the 381 galaxies that were missed by the Bayesian deblending analysis.
We performed a visual inspection of these objects to verify the accuracy of the {\it GALEX} pipeline apertures.
Custom photometry was required for $\sim$15\% of this sample owing primarily to object blends in the pipeline apertures.
The final bright/extended sample consists of 868 galaxies located within the inner 0.6$^{\circ}$ {\it GALEX} FOV.

\subsection{Completeness Limits of the UV Catalogs}
Simulations were performed to assess the completeness of the Bayesian deblending and {\it GALEX} pipeline catalogs.
Point sources are a good approximation for the majority of galaxies in our {\it GALEX} image as
82\% of galaxies in the SDSS catalog have optical half-light radii that are smaller than a {\it GALEX} pixel (1.5$\arcsec$).
A total of $\sim$4000 artificial point sources were randomly distributed across the {\it GALEX} NUV and FUV pipeline images, divided into
six smaller simulations to maintain a source density that is representative of the original image.
Point sources were convolved with the {\it GALEX} PSF and random (Poisson) photon noise was added to simulate real objects.

The source detection rates for the simulated point sources are presented in Figure \ref{deteff}.
The Bayesian deblending algorithm recovered 95\% of the simulated sources to m$_{uv}$$\sim$24.5 in each waveband.
The detection rates using SExtractor are also shown in Figure \ref{deteff}, which was run in the same configuration as the {\it GALEX} pipeline.
SExtractor detection rates are similar to Bayesian deblending for sources brighter than m$_{uv}$=21 in both {\it GALEX} bands.
At fainter magnitudes, SExtractor gives relatively lower detection rates due to object blends and source confusion (discussed in \S3.4).
We note that previous studies with deep {\it GALEX} imaging have used alternative SExtractor parameters in order to improve source detection for faint objects \citep[e.g.][]{Arnouts2005,Xu2005}.
We have tested the SExtractor detection rates for the configuration used in \citet[][available from the online version of their Table 1]{Xu2005}, and found only a
marginal difference with the detection rates for the {\it GALEX} pipeline shown in Figure \ref{deteff}.  Thus we conclude that the Bayesian deblending method
presents advantages over SExtractor even with these proposed improvements to the SExtractor parameters.

As such, we rely on the Bayesian deblending catalog for sources fainter than m$_{uv}$=21 in both bands, and use the {\it GALEX} pipeline catalog for brighter galaxies.
The UV detection efficiency for our final source catalog is higher than the rates shown in Figure \ref{deteff}, as we have manually added objects that were missed by both detection methods
(8 and 381 galaxies were added to the bright pipeline sample and the faint Bayesian deblending sample, respectively). The added detections are roughly one-half of the expected missing sources.
We therefore assume the UV detection efficiency is midway between 100\% completeness and the rates shown in Figure \ref{deteff}, with upper/lower error limits that span this range.


\subsection{Photometry Limits of the UV Catalogs}
Next we evaluated the photometric accuracy of both the {\it GALEX} pipeline and Bayesian deblending catalogs.
In Figure \ref{magdiff} we compare the artificial magnitudes of the simulated point sources to the magnitudes that were measured using both photometry methods.
In each panel we show the average offset and the 1$\sigma$ rms scatter between the simulated and measured magnitudes.
For simulated objects brighter than $NUV$$=$21 ($FUV$=23), both methods of photometry recover the simulated magnitudes to within 0.1 mag.
At fainter magnitudes, Bayesian deblending is less affected by systematic errors as compared to SExtractor, and recovers the simulated magnitudes with good accuracy to $NUV$=24.5 and
$FUV$=25.0 (i.e.~the average offset is nearly zero at these limits with a 1$\sigma$ rms error = 0.3 mag).
Interestingly, the average magnitude offset for the pipeline does not deviate far from zero at similar faint magnitude limits, thus it is still useful for statistical studies (provided incompleteness is properly addressed)
but unreliable for studies of individual galaxies.
We have identified four systematic errors that influence flux measurements in {\it GALEX} images whose importance varies among both methods of photometry and the two {\it GALEX} filters:

\begin{trivlist}
\item [~\bf{Object Blends}]~Object blends result from the inability to separate two or more objects that are in chance alignment on the sky.
Based on the average offset line shown in the top-left panel of Figure \ref{magdiff}, SExtractor NUV magnitudes are 0.05-0.15
mag too bright for sources between 21.0$\simlt$$NUV$$\simlt$23.0, respectively, due to object blends.
The average magnitude offset will likely grow at fainter magnitudes but another systematic effect (source confusion) is dominant at $NUV$$>$23.0.
The influence that object blends have on SExtractor photometry is clear by comparing the average offsets for Figures \ref{magdiff} and \ref{magdiff_ISO}, 
as the latter diagram shows the same magnitude comparison but for isolated sources.
The SExtractor FUV photometry is less affected by object blends owing to fewer galaxy detections as compared to the NUV band (resulting from a lower survey volume due to FUV dropouts),
and a smaller PSF for the FUV filter.
The Bayesian deblending method is relatively less affected by object blends in either band owing to prior knowledge of the source coordinates.

\item [~\bf{Source Confusion}]~Objects fainter than the confusion limit typically have underestimated flux measurements and large scatter due
to the `sea of unresolved sources' at fainter magnitudes that modulate the background level \citep[][]{Hogg2001}.
The confusion limit is a function of the number density of sources and the instrumental resolution, such that images with a relatively steep Log N/Log S
source distribution and/or poor resolution have a brighter confusion limit.
Our simulations indicate that source confusion affects only the SExtractor NUV photometry for our {\it GALEX} image.
This is visible in the top-left panel of Figure \ref{magdiff} (also in Figure \ref{magdiff_ISO}), where the magnitudes are underestimated across
a bump feature on the average offset line between 23$<$$NUV$$<$24.5.
The average offset will likely increase at fainter magnitudes but another systematic effect (the Eddington Bias) affects the SExtractor photometry at $NUV$$\simgt$24.0.
At FUV wavelengths, the SExtractor photometry is not affected by source confusion (at least to $FUV$=24) owing to fewer galaxy detections and a relatively smaller PSF as compared to the NUV band.
The Bayesian deblending catalog is not affected by source confusion in either {\it GALEX} band over the magnitude range studied here.
This is expected since the confusion limit depends on the resolution of the detection image (i.e.~the SDSS image), which has $\sim$4x higher resolution than {\it GALEX}.

\item [~\bf{Eddington Bias}]~The Eddington Bias \citep{Eddington1913} describes a source population with flux measurements that are systematically higher than their true flux values owing to Poisson noise.
It arises across any magnitude interval that samples a rising Log N/Log S source distribution, although its effect is greatest near
the detection limit where a relatively high fraction of sources have overestimated flux measurements.
Every magnitude comparison shown in Figure \ref{magdiff} (or Figure \ref{magdiff_ISO}) is affected at faint
magnitudes by the Eddington Bias, which is responsible for the downturn of the average offset line at faint magnitudes.
The Eddington Bias affects the Bayesian deblending catalog at slightly fainter magnitudes than the SExtractor catalog (and with a more gradual downturn),
which is expected when source detection is performed in the higher signal-to-noise SDSS images \citep[][]{Hogg1998}.
We choose a faint-end magnitude limit of $NUV$=24.5 and $FUV$=25.0 for the Bayesian deblending catalog in order to avoid the Eddington Bias.

\item [~\bf{Aperture Correction}]~The {\it GALEX} PSF extends appreciably beyond the SExtractor Kron apertures, such that 
SExtractor measurements are underestimated by $\sim$0.1 dex in the NUV band and $\sim$0.07 dex in FUV across all simulated magnitudes.
We have corrected the SExtractor magnitudes in Figures \ref{magdiff} and \ref{magdiff_ISO} for this nearly constant offset in order to isolate the effects discussed above.
Bayesian deblending is not affected by light loss since the light distribution is fit across the entire {\it GALEX} PSF profile.
\end{trivlist}

Although Bayesian deblending provides reliable photometry for unresolved objects, we expect that its flux measurements are underestimated
for galaxies with apparent sizes that are much larger than the {\it GALEX} PSF.
In Figure \ref{magdiff_ext} we compare the magnitude difference for galaxies in the Bayesian deblending and {\it GALEX} pipeline catalogs versus
the size of the galaxy (taken as the diameter that encloses 90\% of the SDSS $r$-band flux).
The diagram confirms that the magnitudes diverge at a galaxy size of $\sim$10$\arcsec$.
As such, we adopt the photometry in the {\it GALEX} pipeline catalog for more extended galaxies as it provides a better estimate of the total galaxy flux.

It is well known that the SExtractor Kron photometry has a missing light problem, as
no finite aperture can capture 100\% of the light distribution for a galaxy that follows a S\'ersic profile \citep[even for the case of infinite S/N;][]{Graham2005}.
In addition, the fraction of missing light is larger for detections with relatively low S/N (e.g.~low surface brightness (LSB) galaxies) 
as the Kron aperture is scaled only from pixels that satisfy the detection threshold.
We estimate that this effect is relatively small for galaxies in the Coma cluster based on the procedure
for estimating the missing SExtractor flux described in \cite{Graham2005}.
For example, the majority of Coma member galaxies are only marginally resolved in the {\it GALEX} images,
for which SExtractor adopted its minimum Kron radius value of 3.5 pixels (which corresponds to $\sim$2.5 kpc at the distance of Coma).
Assuming a S\'ersic index of n=1 for dwarf galaxies and normal spirals (i.e.~the majority of our confirmed Coma members) 
and effective radii between $\sim$1-2.5 kpc \citep[e.g.][]{Graham2003b},
we estimate that the SExtractor Kron apertures are missing $\sim$5-10\% of the source flux (0.05-0.1 mag).
This is the same amount of missing flux we estimated for the SExtractor photometry of point sources based on our simulations.
As such, we have applied a small correction to the SExtractor Kron magnitudes ($\Delta$$m$=0.1 $\pm$0.05 mag) in both {\it GALEX} bands in order
to better recover the total galaxy magnitude. Note that the same analysis suggests that the SExtractor photometry for giant early-type galaxies in the
Coma cluster with a S\'ersic index n=4 (i.e.~a de Vaucouleur profile) may be missing $\sim$25\% of the light distribution, thus are still too faint by $\sim$0.2 dex in magnitude 
after performing corrections.

\subsection{Summary of the Final {\it GALEX}/SDSS Catalog}
We have assembled a final source catalog that consists of 9700 objects with {\it GALEX} and SDSS photometry.
UV photometry is taken from the {\it GALEX} pipeline catalog for bright or extended galaxies,
i.e.~objects brighter than $NUV$=21 or $FUV$=21, or with an optical 90\% light diameter larger than 10$\arcsec$ in the SDSS $r$-band.
We rely on a Bayesian deblending method for faint compact objects with magnitudes between 21$<$$NUV$$<$24.5 and 21$<$$FUV$$<$25.
The catalog includes all such objects located within the inner 0.5 deg {\it GALEX} FOV (0.8 deg$^{2}$), 
and provides slightly wider coverage (the inner 0.6 deg {\it GALEX} FOV or 1.1 deg$^{2}$) for galaxies brighter than $NUV$($FUV$)$\leq$21.
The optical counterparts to the UV sources are either (a) brighter than $r$=24 and classified as galaxies by SDSS (8704 sources), or (b) have magnitudes between 21$<$$r$$\leq$24 and
are classified as stars by SDSS (996 sources). For the latter case, we expect that over one-half of these objects are actually galaxies based on a comparison to HST-ACS data.
A subset of our {\it GALEX}/SDSS catalog is presented in Table 1 (the full table is available in the electronic version of this paper). For each source we list the SDSS coordinates [1-2],
SDSS `type' (galaxy=3, star=6) [3], SDSS Petrosian magnitudes [4-8], and the {\it GALEX} NUV and FUV magnitudes including uncertainties [9-12].

The UV completeness limit of our {\it GALEX}/SDSS catalog is a function of both the {\it GALEX} and SDSS detection limits.
Magnitude histograms provide a straightforward diagnostic of the completeness limits for multi-wavelength catalogs.
In Figure \ref{uvhisto} we present the UV magnitude histograms for all sources in our catalog separated by $r$-band magnitude.
The diagram indicates that our catalog includes the majority of galaxies at the adopted limits of $NUV$=23.0 and $FUV$=23.5,
which were chosen at $\sim$0.5 mag brighter than the observed flattening and then decline in the number of UV/optical detections.
At fainter UV magnitudes, the number of detections is limited by the SDSS incompleteness.

It is also useful to describe the completeness of our {\it GALEX}/SDSS catalog relative to optical magnitudes.
In Figure \ref{uvfrac} we show the fraction of sources in the SDSS DR6 that are members of the our catalog.
The diagram shows that our catalog includes $\sim$95\% of all SDSS DR6 galaxies brighter than $r$=18.5 ($r$=17.5) for the NUV (FUV) filters, respectively.
The SDSS fraction drops at fainter optical magnitudes as we lose coverage of red quiescent galaxies (e.g.~$NUV$-$r$=6.0 and $FUV$-$r$=7.5),
which are fainter than the detection/photometry limit of our {\it GALEX} image ($NUV$$_{lim}$=24.5 and $FUV$$_{lim}$=25.0).
The dependence of our completeness limits on UV-optical color is more obvious from the color magnitude diagrams presented in Figure \ref{colormag} ($NUV$-$r$ and $FUV$-$r$ vs $r$).
Galaxies with UV-optical colors bluer than $NUV$-$r$=2.3 and $FUV$-$r$=2.8 are limited
by the SDSS completeness limit ($r$=22.2) as opposed to the depth of our {\it GALEX} image.

Galaxies that are spectroscopically-confirmed members of the Coma cluster are also shown in Figure \ref{colormag}.
The redshifts were taken from several spectroscopic surveys \citep[e.g.~R.~Marzke et al.~2010, {\it in prep};][]{Adelman2008,Mobasher2001,Colless1996}, which are described in our 
companion paper (Paper II; D.~Hammer et al.~2010, {\it submitted}).
The cluster red sequence is the dense band of Coma members that stretches horizontally across the top of each diagram.
From the NUV CMD, the red sequence traces bluer colors at faint optical magnitudes until we lose coverage at $r$$\sim$20, i.e.~NUV detections are available for
all Coma members brighter than M$_{r}$=-15.
The red sequence has a relatively flat slope in the FUV CMD, thus we lose coverage of red sequence galaxies at
a brighter optical magnitude as compared to the NUV band ($r$$\sim$17.5 or M$_{r}$=-17.5).
The red sequence detection limits for both {\it GALEX} bands are in the optical magnitude range expected for dwarf early-type (dE) galaxies.

\section{Comparison to Previous UV Surveys}
Although our {\it GALEX} image covers part of the Coma cluster ($z$$\sim$0.023), its depth allows us to constrain the background galaxy population as
cluster member galaxies account for only a few percent of all UV-detected galaxies at the limiting depth of our {\it GALEX}/SDSS catalog.
This allows for a comparison of the UV galaxy counts measured using our catalog as compared to early {\it GALEX} measurements
that relied on the {\it GALEX} pipeline catalog (i.e.~before improved source detection and photometry techniques, such as Bayesian deblending, were introduced).
In the following sections we re-measure the number counts of galaxies to faint UV magnitudes, and use these values to
estimate the source confusion limits for both the Bayesian deblending catalog and the {\it GALEX} pipeline catalogs.

\subsection{Galaxy Number Counts}
Measurements of the UV number counts of galaxies are important for constraining models of galaxy formation and evolution \citep[e.g.~semi-analytic $\Lambda$CDM models;][]{Nagashima2002,Gilmore2009}.
The differential number counts for our {\it GALEX} survey are shown in Figure \ref{uvdiff}, which covers the magnitude range 17.0-23.5 for the FUV filter and 16.5-23.0 for the NUV filter.
We derived the galaxy counts by counting objects in our catalog that are considered galaxies by SDSS, and then performing statistical corrections to account for
(a) misclassified stars in the SDSS catalog (see Appendix A), (b) the Eddington Bias, which was estimated following the formalism in \cite{Eddington1913},
and (c) UV incompleteness based on our simulations.
The statistical corrections account for $\simlt$10\% (5\%) of the galaxy counts across the magnitude range covered by the FUV (NUV) bands, respectively.
The error bars include Poisson counting errors \citep[][]{Gehrels1986}, uncertainties for the SDSS star/galaxy classification (see Appendix A), errors for the
Eddington Bias correction (conservatively taken as 50\% of the correction), and the uncertainty of our UV detection efficiency from simulations.
The error bars also include cosmic variance, which we estimated following the formalism in \cite{Glazebrook1994} using typical parameters for the galaxy correlation function at UV wavelengths
\citep[$\gamma$=1.6; r$_{0}$=3.0/4.0 Mpc h$^{-1}$ for galaxies brighter/fainter than FUV(NUV)=21, respectively;][]{Milliard2007,Heinis2009}.
In Table 2 we list our differential number counts and the upper/lower limits [2-4], the raw number of galaxies in each magnitude bin [5], and the survey area [6].

Galaxy counts from previous UV surveys are shown in Figure \ref{uvdiff} for comparison, such as a relatively shallow survey performed with the balloon-borne FOCA instrument at 2000 \AA~\citep{Milliard1992},
a {\it GALEX} study of multiple MIS and DIS fields \citep{Xu2005}, a Swift-UVOT (uvm2) NUV survey of the CDF-S \citep{Hoversten2009}, a deep HST-STIS (F25QTZ) FUV and NUV survey that observed the HDF-N and HDF-S fields
\citep[][]{Gardner2000}, and a deep HST-ACS (SBC F150LP) FUV survey that also covers the HDF-N field but with 3x the coverage area \citep[][]{Teplitz2006}.
We also include a model for the galaxy counts in the {\it GALEX} FUV and NUV bands \citep{Xu2005}.
We have performed a color correction to match the FOCA filter to the {\it GALEX} system based on a magnitude comparison for 46 galaxies in our {\it GALEX} FOV that were also detected by FOCA \citep[][]{Donas1991}.
We converted the FOCA monochromatic magnitudes at 2000 \AA~using the relations $FUV$=$m_{2000}$+2.4 and $NUV$=$m_{2000}$+2.0, which are reliable for star forming galaxies brighter than m$_{2000}$=17.5;
we found that FOCA significantly overestimates the flux for galaxies fainter than m$_{2000}$=17.5 as suggested in previous {\it GALEX} studies \citep[e.g.][]{Wyder2005}.
We do not perform color corrections for the STIS NUV and UVOT uvm2 filters, as their similarity with the {\it GALEX} NUV bandpass results in a negligible $\sim$0.02 mag correction across all three filters \citep{Hoversten2009}.
Color corrections for the HST FUV filters are beyond the scope of this paper, as they are highly sensitive to redshift \citep[e.g.][]{Teplitz2006}.

We have also plotted in Figure \ref{uvdiff} the expected number of Coma members in each apparent magnitude bin based on measurements of the Coma UV luminosity function (Paper II).
As expected, our total number counts are higher than the previous {\it GALEX} measurements at bright magnitudes \citep[$NUV$ and $FUV$$\simlt$21;][]{Xu2005}
owing to the overdensity of galaxies in the foreground Coma cluster.
However, we see a significant excess of faint galaxies relative to the \cite{Xu2005} counts at $NUV$=21.5-23 and $FUV$=21.5-23.5, amounting to 50\% in NUV and 60\% in the FUV counts at our limiting depth.
This is more obvious from Figure \ref{uvdiff_baseline}, where we have plotted galaxy counts relative to an arbitrary baseline (the model counts).
The excess galaxy counts cannot be accounted for by Coma members, which contribute only 2\% of the counts at the limiting depth.
In addition, we suspect that our counts are too low across the three faintest magnitude bins in both {\it GALEX} bands owing to SDSS incompleteness.
For instance, the SDSS detection efficiency for extended galaxies may only be 50\% at $r$=22.2, while it is 95\% for point sources at this magnitude limit\footnote[9]{www.sdss.org/dr6/products/general/completeness.html}.
Assuming a worst-case scenario such that one-half of SDSS galaxies at $r$$>$21 are missing from our catalog,
our UV counts could be underestimated by $\sim$5, 10, and 30\% across our three faintest magnitude bins for both {\it GALEX} bands.

Our excess galaxy counts and their slopes at faint magnitudes are a better match to the deep HST surveys.
At NUV wavelengths, our counts are statistically consistent with the deeper HST-STIS counts where the two datasets are bridged;
the large STIS uncertainties, however, prevent any definitive conclusions regarding this agreement as we could also argue that our counts are lower by a factor of 2.
Our FUV counts are clearly disjoint with both HST surveys, such that the deep HST FUV counts are 30-50\% larger than expected based on an extrapolation of our values to fainter magnitudes.
This difference may result from incompleteness of the SDSS catalog as discussed above.
Interestingly, \cite{Teplitz2006} suggest that the HST-SBC FUV survey may cover a 30\% larger volume than {\it GALEX} owing to its slightly redder wavelength coverage (the HST-STIS and SBC FUV filters
extend $\sim$100 and 150 \AA~redward of the {\it GALEX} FUV filter, respectively).
A more recent HST-SBC FUV study covering a larger area with more sightlines (E.~Voyer et al.~2010, {\it in prep}) finds $\sim$30\% lower counts
(on average) than the \cite{Teplitz2006} or \cite{Gardner2000} studies, possibly due to cosmic variance.
A combination of these effects is likely responsible for the offset between our FUV counts and the deeper HST surveys.

The excess galaxy counts are not likely attributed to the different detection/photometry methods and statistical corrections used in our study as compared to \cite{Xu2005}.
For example, we measured nearly identical FUV and NUV counts using the {\it GALEX} pipeline catalog with the exact selection criteria and corrections used by \cite{Xu2005}.
The excess galaxy counts are at least partially due to AGN that are not accounted for in this study, but were removed by \cite{Xu2005}.
However, only a small fraction of galaxies are expected to have UV emission that is dominated by AGN \citep[e.g.~a few percent;][]{Hoversten2009}.
We also note that the UVOT NUV counts are approximately midway between both sets of {\it GALEX} measurements, but we suspect that their true {\it GALEX}-matched values are closer to our counts,
e.g.~the redder wavelength coverage of the {\it GALEX} NUV filter relative to UVOT uvm2 results in both more detections of early-type galaxies and a slightly larger survey volume
as the Lyman limit passes through the {\it GALEX} NUV band at a higher redshift.
An excess of background galaxies owing to cosmic variance would seem unlikely, as this requires a minimum 5$\sigma$ overdensity in the background Coma field at our limiting depth, 
where the cosmic variance is $\sim$10\% (the excess is $\sim$50-60\% relative to \cite{Xu2005}).
For example, background clusters are not likely responsible for the excess counts as this would require at least 25 Coma cluster galaxy populations in the background FOV.
However, massive clusters such as Coma tend to be associated with more filaments \citep[e.g.][]{Pimbblet2004}, and we cannot rule out excess galaxy counts due to such large-scale structures.
Interestingly, \cite{Jenkins2007} also reported excess galaxy counts in the same Coma-3 region at 3.6 $\micron$ relative to a comparison field.
Clearly, further studies are needed to explain the difference between both sets of {\it GALEX} measurements.


In each diagram we have plotted the UV galaxy count models presented in \cite{Xu2005}. These models were computed assuming pure luminosity evolution
($L_{*}$$\sim$(1+$z$)$^{2.5}$) using a single starburst SED \citep[SB4;][]{Kinney1996} with a flat spectrum blueward of Ly$\alpha$
and no emission below the Lyman limit.
Although the models assume a relatively simple evolution and only allow for a single SED, they provide a good first-order fit to both the FUV and NUV galaxy counts across $\sim$12 magnitudes.
Noticeable shortcomings are that (a) whereas the models previously overestimated the faint galaxy counts from early {\it GALEX} measurements, they are now too low relative to our galaxy counts for both {\it GALEX} bands,
(b) the slope of the FUV model flattens at FUV=22, which occurs $\sim$2-3 mags too bright as compared to the measured FUV counts, and 
(c) the slope of the FUV and NUV models are too steep at very faint magnitudes ($NUV$ and $FUV$$\simgt$25).
The difference between the model and measured counts likely results from neglecting number density evolution, using a single SED to represent all galaxies, assuming a flat spectrum shortward of Ly-$\alpha$, 
and poor constraints on the faint-end slopes of restframe UV LFs at higher redshifts ($z$$\simgt$0.5).
For example, cosmic downsizing \citep{Cowie1996} predicts some number density evolution via low-mass galaxies that ``turn on'' at more recent epochs.
Also, at higher redshifts ($z$$\simgt$0.5) the majority of UV-detected galaxies are unobscured starbursts \citep[e.g.~Kinney SB1 template;][]{Arnouts2005},
as opposed to the obscured starburst template used for the models.
The effect that these parameters have on the UV galaxy counts is highly model-dependent, and thus beyond the scope of this paper.
Our counts will allow for more improved models of galaxy evolution by constraining number counts at intermediate UV magnitudes.

Finally, although the FOCA counts were measured for the field galaxy population, there is good agreement between FOCA and our measurements at bright magnitudes ($NUV$ and $FUV$$<$19.5).
This agreement may result from contamination by cluster galaxies in the FOCA survey field SA57, which is partially located inside the apparent virial radius of the Coma cluster.
Although the FOCA counts were measured in three separate fields, the majority of bright galaxies were detected in the SA57 field.
At fainter magnitudes, the FOCA counts are higher than all other studies likely owing to overestimated flux measurements.

\subsection{Source Confusion Limits}
We have already established that our UV photometry is not affected by source confusion over the magnitude range studied here.
This results from the use of a Bayesian deblending technique that performs source detection in higher-resolution SDSS images and then photometry in the {\it GALEX} image.
Source confusion affects the {\it GALEX} pipeline catalog at relatively brighter magnitudes owing to source detection and photometry that are performed in the same {\it GALEX} image.
Given the higher galaxy counts reported in this study, it is necessary to revise estimates of the source confusion limit in {\it GALEX} pipeline catalogs.

Although studies typically use indirect methods to measure the source confusion limit (discussed in the next paragraph), the most reliable estimates are performed via simulations \citep{Hogg2001}.
Our simulations indicate that the {\it GALEX} pipeline catalogs are confusion limited in the NUV band between $NUV$=23-23.5 based on the abrupt drop in the detection efficiency and systematic
photometry errors that emerge over this magnitude range (\S 3.4).
Our FUV image is flux limited, thus simulations were only able to establish an upper (bright) magnitude limit of $FUV$=24 where the {\it GALEX} pipeline is not affected by source confusion.

In lieu of performing simulations, the source confusion limit may be estimated indirectly from the measured number density of sources and the instrument beam size,
more commonly referred to as the number of beams-per-source ($b/s$):
\begin{equation}
b/s = n(<m)^{-1}\,\Omega_{beam}^{-1},
\label{confusion}
\end{equation}
where $n($$<$$m)$ is the cumulative number counts at apparent magnitude $m$, and $\Omega_{beam}$ is the instrument beam size.
\cite{Hogg2001} suggests that $b/s$=40 should be taken as the confusion limit for source distributions with a cumulative Log N-Log S slope less than 1.5; 
we adopt this value as our cumulative number counts have slopes of $\beta$$_{nuv}$=1.45 and $\beta$$_{fuv}$=1.25, respectively.
The cumulative galaxy counts for our survey are presented in Figure \ref{uvdiff_cumul}.
One complication associated with equation [\ref{confusion}] is that there does not exist a standard definition for the beam size,
e.g.~the beam size may be taken as the diameter equal to (a) the 1-$\sigma$ Gaussian profile ($\simeq$FWHM/2.35), (b) the FWHM, or (c) the measured 50\% light diameter.
We chose a beam size equal to the {\it GALEX} FWHM, as only this definition of the beam size corresponds to a 40 beams-per-source NUV confusion limit ($NUV$=23.1) that lies within the
magnitude range $NUV$=23.0-23.5 predicted by our simulations.
Applying the {\it GALEX} FUV FWHM (4.8$\arcsec$) and our FUV number counts to equation [\ref{confusion}], we estimate that the {\it GALEX} pipeline catalogs are confusion limited at $FUV$=24.0.
We conclude that source confusion affects the {\it GALEX} pipeline catalogs at magnitudes fainter than $NUV$=23 and $FUV$=24, which are approximately $\sim$1 mag brighter 
than the source confusion limits estimated by \cite{Xu2005}.

The Bayesian deblending method is affected by source confusion at relatively fainter magnitudes than the {\it GALEX} pipeline catalog owing to source detection that is performed in 
higher-resolution optical images (SDSS in our case).  Adopting the standard 1.4$\arcsec$ SDSS seeing for the beam size in equation [\ref{confusion}], we estimate that our Bayesian deblending method 
would become confusion limited at $NUV$$\sim$25.5 and $FUV$$\sim$27.

\section{Future Work}
We use the {\it GALEX}/SDSS source catalog presented in this paper and a deep database of spectroscopic redshifts to measure the UV luminosity function (LF) of the Coma cluster (Paper II; D.~Hammer et al.~2010, {\it submitted}).
The improved source detection and photometry via the Bayesian deblending technique allows us to measure the deepest UV LF presented for a cluster thus far.
We were recently awarded a deep {\it GALEX} observation at the center of the Coma cluster (P.I.~R. Smith) that will allow for a comparison of the UV-faint properties of Coma member
 galaxies across a large range of cluster-centric distance.

\acknowledgements
We thank Antara Basu-Zych for commenting on a draft of this paper,
Elysse Voyer for sharing new results, Panayiotis Tzanavaris and Bret Lehmer for useful science discussion, and Eric Cardiff for providing helpful language translations.
This research was supported by the {\it GALEX} Cycle 2 grant 05-GALEX05-0046 (P.I.~Hornschemeier). {\it GALEX} is a NASA Small Explorer, developed 
in cooperation with the Centre National d'Etudes Spatiales of France and the Korean Ministry of Science and Technology. Funding for the creation and 
distribution of the SDSS Archive has been provided by the Alfred P. Sloan Foundation, the Participating Institutions, NASA, the NSF, DoE, Monbukagakusho, 
Max Planck Society, and the Higher Education Funding Council for England. The SDSS Web Site is http://www.sdss.org/. This study made use of the NASA
Extragalactic Database (NED) which is operated by the Jet Propulsion Laboratory (Cal Tech), under contract with NASA.
\clearpage

\appendix
\section{Appendix A: Constraining Properties of SDSS DR6 Sources with HST-ACS Data}
The Coma cluster was the target of a HST-ACS Treasury survey that provides coverage in 19 fields located at the center of Coma with six additional 
fields in the south-west region of the cluster \citep{Carter2008}.
The entire ACS footprint is covered by the SDSS DR6, and the six ACS fields in the south-west region of Coma overlap our {\it GALEX} FOV.
Although the ACS coverage of our {\it GALEX} field is negligible (a few percent of the total {\it GALEX} FOV), we have used the ACS data 
to assess the reliability of the SDSS catalog, which we rely on for galaxy identification in this study.
Specifically, the ACS images have relatively high resolution \citep[0.12$\arcsec$;][]{Ford1998} and sensitivity (95\% complete to $r$$\sim$26; D.~Hammer et al.~2010, {\it in prep}) as compared to SDSS,
which allows us to determine statistically (a) the fraction of real/spurious objects in SDSS DR6 at faint magnitudes, and (b) the accuracy of the SDSS star/galaxy classification at faint magnitudes.
\begin{trivlist}
\item [~\bf{Real/Spurious Detections in the SDSS DR6}]~We have estimated the fraction of real/spurious detections in the SDSS DR6 catalog 
by searching the ACS images at the location of every SDSS object in the ACS footprint.
The results are given in Table 3, which lists the SDSS $r$-band magnitude [1], the number of SDSS galaxies inspected [2],
the fraction of real objects among SDSS galaxies and the 1$\sigma$ upper/lower confidence limits [3-5], the number of SDSS stars inspected [6], and the fraction of real objects among SDSS stars
and the 1$\sigma$ upper/lower confidence limits [7-9]. The uncertainties were calculated following the method in \cite{Gehrels1986} for binomial statistics.
We found that the fraction of real objects among SDSS galaxies/stars is $\simgt$85\% at magnitudes brighter than $r$=24 mag.
At fainter magnitudes, 40\% of objects listed in the SDSS DR6 are spurious detections that result primarily from galaxy shredding near bright galaxies.
\\
\item [~\bf{Accuracy of the SDSS Star/Galaxy Classification}]~The SDSS pipeline reports that its star/galaxy (S/G) classification is 95\% accurate to $r$=21 \citep[e.g.][]{Stoughton2002},
although past studies suggest it may perform well to `at least $r$=21.5' \citep{Lupton2001}.
We have tested the reliability of the SDSS S/G classification at faint magnitudes by comparing it to morphologies taken directly from the ACS images.
This analysis was performed for SDSS objects located inside the ACS footprint, and also for the subset of SDSS/ACS objects that were detected
in our {\it GALEX} field (for objects brighter than our UV 95\% completeness limits of $NUV$=24.5 and $FUV$=25.0).
The results are shown in Table 4 which lists the SDSS $r$-band magnitude [1], the number of SDSS-classified galaxies that were inspected [2],
the fraction of true galaxies among SDSS galaxies and the 1$\sigma$ upper/lower confidence limits [3-5], the number of SDSS-classified stars that were inspected [6],
and the fraction of true galaxies among SDSS stars and the 1$\sigma$ upper/lower confidence limits [7-9]. The uncertainties were calculated following the method in \cite{Gehrels1986} for binomial statistics.

For SDSS detections alone, we confirm that the SDSS S/G classification is $\sim$95\% accurate for all objects brighter than $r$=21.
At fainter magnitudes, objects that are classified as galaxies by SDSS remain $\simgt$90\% accurate to the faintest limits of the SDSS survey ($r$$\sim$25).
On the other hand, SDSS star classifications are unreliable at magnitudes fainter than $r$=21, e.g.~over one-half of faint ($r$$>$21) objects considered stars by SDSS are actually galaxies.

For the subset of SDSS sources with {\it GALEX} FUV or NUV detections, objects considered galaxies by SDSS are $\simgt$95\% accurate to $r$=25.0, which is a slightly higher rate as compared to SDSS detections alone.
There are relatively few SDSS-classified stars with {\it GALEX} detections that are located inside the ACS footprint (12 FUV detections and 31 NUV detections).
The subset of SDSS stars that are brighter than $r$=21 were indeed confirmed as stars (8/8).
On the other hand, all SDSS star classifications fainter than $r$=21 were confirmed as galaxies for {\it GALEX} FUV detections (11/11), and also for the majority of NUV detections (18/24);
these rates are significantly higher than for SDSS detections alone, which suggests that the UV data tends to select galaxies that were misclassified by SDSS.
This is likely a color selection effect owing to optical-selected stars that typically have red UV-optical colors (thus are fainter than the UV completeness limits of our catalog),
while the background galaxies tend to be spirals with blue UV-optical colors.

Statistical studies that use our {\it GALEX}/SDSS catalog (e.g.~UV galaxy counts, UV luminosity function) must correct for the number of galaxies among objects classified as stars by SDSS at magnitudes fainter than $r$=21.0.
For NUV studies, this correction may be estimated by counting the number of SDSS stars in each UV magnitude bin, then separating the stars into $r$-band bins, and then applying to the corrections listed in the middle panel of Table 4,
e.g.~the NUV correction for misclassified galaxies is: $g$$_{nuv}$=0.50$^{0.31}_{0.31}$ (21$<$$r$$<$22), $g$$_{nuv}$=0.77$^{0.12}_{0.18}$ (21$<$$r$$<$22), $g$$_{nuv}$=0.86$^{0.12}_{0.26}$ (23$<$$r$$<$24).
At FUV wavelengths, the fraction of galaxies among SDSS classified stars is relatively higher as compared to the NUV band (stars are less likely to be observed in the FUV band),
although the 100\% galaxy correction listed in the bottom panel of Table 4 is likely too high owing to low-number statistics.
Instead, we recommend using a FUV galaxy correction that is midway between the NUV fraction given above and a 100\% correction, with uncertainties that span the total range,
e.g.~the FUV correction for misclassified galaxies is: $g$$_{fuv}$=0.75$^{0.25}_{0.25}$ (21$<$$r$$<$22), $g$$_{fuv}$=0.88$^{0.12}_{0.12}$ (21$<$$r$$<$22), $g$$_{fuv}$=0.93$^{0.07}_{0.07}$ (23$<$$r$$<$24).
\end{trivlist}

\clearpage
\bibliographystyle{apj}
\bibliography{ms}

\begin{thebibliography}{}

\bibitem[\protect\citeauthoryear{{Abazajian} et~al.}{{Abazajian}
  et~al.}{2004}]{Abaz2004}
{Abazajian}, K., et~al. 2004, \aj, 128, 502

\bibitem[\protect\citeauthoryear{{Abell}, {Corwin}, \& {Olowin}}{{Abell}
  et~al.}{1989}]{Abell1989}
{Abell}, G.~O., {Corwin}, H.~G., Jr.,  \& {Olowin}, R.~P. 1989, \apjs, 70, 1

\bibitem[\protect\citeauthoryear{{Adelman-McCarthy} et~al.}{{Adelman-McCarthy}
  et~al.}{2008}]{Adelman2008}
{Adelman-McCarthy}, J.~K., et~al. 2008, \apjs, 175, 297

\bibitem[\protect\citeauthoryear{{Andreon}}{{Andreon}}{1999}]{Andreon1999}
{Andreon}, S. 1999, \aap, 351, 65

\bibitem[\protect\citeauthoryear{{Arnouts} et~al.}{{Arnouts}
  et~al.}{2005}]{Arnouts2005}
{Arnouts}, S., et~al. 2005, \apjl, 619, L43

\bibitem[\protect\citeauthoryear{{Bautz} \& {Morgan}}{{Bautz} \&
  {Morgan}}{1970}]{Bautz1970}
{Bautz}, L.~P.,  \& {Morgan}, W.~W. 1970, \apjl, 162, L149

\bibitem[\protect\citeauthoryear{{Bertin} \& {Arnouts}}{{Bertin} \&
  {Arnouts}}{1996}]{Bertin96}
{Bertin}, E.,  \& {Arnouts}, S. 1996, \aaps, 117, 393

\bibitem[\protect\citeauthoryear{{Blanton} \& {Roweis}}{{Blanton} \&
  {Roweis}}{2007}]{Blanton2007}
{Blanton}, M.~R.,  \& {Roweis}, S. 2007, \aj, 133, 734

\bibitem[\protect\citeauthoryear{{Boselli} et~al.}{{Boselli}
  et~al.}{2005}]{Boselli2005b}
{Boselli}, A., et~al. 2005, \apjl, 629, L29

\bibitem[\protect\citeauthoryear{{Briel} et~al.}{{Briel}
  et~al.}{2001}]{Briel2001}
{Briel}, U.~G., et~al. 2001, \aap, 365, L60

\bibitem[\protect\citeauthoryear{{Brown}, {Ferguson}, \& {Davidsen}}{{Brown}
  et~al.}{1995}]{Brown1995}
{Brown}, T.~M., {Ferguson}, H.~C.,  \& {Davidsen}, A.~F. 1995, \apjl, 454, L15

\bibitem[\protect\citeauthoryear{{Budav{\'a}ri} et~al.}{{Budav{\'a}ri}
  et~al.}{2009}]{Budavari2009}
{Budav{\'a}ri}, T., et~al. 2009, \apj, 694, 1281

\bibitem[\protect\citeauthoryear{{Caldwell} \& {Rose}}{{Caldwell} \&
  {Rose}}{1998}]{Caldwell1998}
{Caldwell}, N.,  \& {Rose}, J.~A. 1998, \aj, 115, 1423

\bibitem[\protect\citeauthoryear{{Caldwell} et~al.}{{Caldwell}
  et~al.}{1993}]{Caldwell1993}
{Caldwell}, N., {Rose}, J.~A., {Sharples}, R.~M., {Ellis}, R.~S.,  \& {Bower},
  R.~G. 1993, \aj, 106, 473

\bibitem[\protect\citeauthoryear{{Carter} et~al.}{{Carter}
  et~al.}{2008}]{Carter2008}
{Carter}, D., et~al. 2008, \apjs, 176, 424

\bibitem[\protect\citeauthoryear{{Code}}{{Code}}{1969}]{Code1969}
{Code}, A.~D. 1969, \pasp, 81, 475

\bibitem[\protect\citeauthoryear{{Colless} \& {Dunn}}{{Colless} \&
  {Dunn}}{1996}]{Colless1996}
{Colless}, M.,  \& {Dunn}, A.~M. 1996, \apj, 458, 435

\bibitem[\protect\citeauthoryear{{Cortese} et~al.}{{Cortese}
  et~al.}{2005}]{Cortese2005}
{Cortese}, L., et~al. 2005, \apjl, 623, L17

\bibitem[\protect\citeauthoryear{{Cortese}, {Gavazzi}, \& {Boselli}}{{Cortese}
  et~al.}{2008}]{Cortese2008}
{Cortese}, L., {Gavazzi}, G.,  \& {Boselli}, A. 2008, \mnras, 390, 1282

\bibitem[\protect\citeauthoryear{{Cortese} et~al.}{{Cortese}
  et~al.}{2003a}]{Cortese2003a}
{Cortese}, L., {Gavazzi}, G., {Boselli}, A., {Iglesias-Paramo}, J., {Donas},
  J.,  \& {Milliard}, B. 2003a, \aap, 410, L25

\bibitem[\protect\citeauthoryear{{Cortese} et~al.}{{Cortese}
  et~al.}{2003b}]{Cortese2003b}
{Cortese}, L., {Gavazzi}, G., {Iglesias-Paramo}, J., {Boselli}, A.,  \&
  {Carrasco}, L. 2003b, \aap, 401, 471

\bibitem[\protect\citeauthoryear{{Cowie} et~al.}{{Cowie}
  et~al.}{1996}]{Cowie1996}
{Cowie}, L.~L., {Songaila}, A., {Hu}, E.~M.,  \& {Cohen}, J.~G. 1996, \aj, 112,
  839

\bibitem[\protect\citeauthoryear{{Donas}, {Milliard}, \& {Laget}}{{Donas}
  et~al.}{1991}]{Donas1991}
{Donas}, J., {Milliard}, B.,  \& {Laget}, M. 1991, \aap, 252, 487

\bibitem[\protect\citeauthoryear{{Dorman}, {O'Connell}, \& {Rood}}{{Dorman}
  et~al.}{1995}]{Dorman1995}
{Dorman}, B., {O'Connell}, R.~W.,  \& {Rood}, R.~T. 1995, \apj, 442, 105

\bibitem[\protect\citeauthoryear{{Dorman}, {O'Connell}, \& {Rood}}{{Dorman}
  et~al.}{2003}]{Dorman2003}
{Dorman}, B., {O'Connell}, R.~W.,  \& {Rood}, R.~T. 2003, \apj, 591, 878

\bibitem[\protect\citeauthoryear{{Eddington}}{{Eddington}}{1913}]{Eddington191%
3}
{Eddington}, A.~S. 1913, \mnras, 73, 359

\bibitem[\protect\citeauthoryear{{Ford} et~al.}{{Ford} et~al.}{1998}]{Ford1998}
{Ford}, H.~C., et~al. 1998, in SPIE Conference Series, ed. P.~Y. {Bely} \&
  J.~B. {Breckinridge}, Vol. 3356, 234

\bibitem[\protect\citeauthoryear{{Gardner}, {Brown}, \& {Ferguson}}{{Gardner}
  et~al.}{2000}]{Gardner2000}
{Gardner}, J.~P., {Brown}, T.~M.,  \& {Ferguson}, H.~C. 2000, \apjl, 542, L79

\bibitem[\protect\citeauthoryear{{Gehrels}}{{Gehrels}}{1986}]{Gehrels1986}
{Gehrels}, N. 1986, \apj, 303, 336

\bibitem[\protect\citeauthoryear{{Gilmore} et~al.}{{Gilmore}
  et~al.}{2009}]{Gilmore2009}
{Gilmore}, R.~C., {Madau}, P., {Primack}, J.~R., {Somerville}, R.~S.,  \&
  {Haardt}, F. 2009, \mnras, 399, 1694

\bibitem[\protect\citeauthoryear{{Glazebrook} et~al.}{{Glazebrook}
  et~al.}{1994}]{Glazebrook1994}
{Glazebrook}, K., {Peacock}, J.~A., {Collins}, C.~A.,  \& {Miller}, L. 1994,
  \mnras, 266, 65

\bibitem[\protect\citeauthoryear{{Graham} \& {Driver}}{{Graham} \&
  {Driver}}{2005}]{Graham2005}
{Graham}, A.~W.,  \& {Driver}, S.~P. 2005, Publications of the Astronomical
  Society of Australia, 22, 118

\bibitem[\protect\citeauthoryear{{Graham} \& {Guzm{\'a}n}}{{Graham} \&
  {Guzm{\'a}n}}{2003}]{Graham2003b}
{Graham}, A.~W.,  \& {Guzm{\'a}n}, R. 2003, \aj, 125, 2936

\bibitem[\protect\citeauthoryear{{Greggio} \& {Renzini}}{{Greggio} \&
  {Renzini}}{1990}]{Greggio1990}
{Greggio}, L.,  \& {Renzini}, A. 1990, \apj, 364, 35

\bibitem[\protect\citeauthoryear{{Guillaume} et~al.}{{Guillaume}
  et~al.}{2006}]{Guillaume2006}
{Guillaume}, M., {Llebaria}, A., {Aymeric}, D., {Arnouts}, S.,  \& {Milliard},
  B. 2006, in SPIE Conference Series, ed. E.~R. {Dougherty}, J.~T. {Astola},
  K.~O. {Egiazarian}, N.~M. {Nasrabadi}, \& S.~A. {Rizvi}, Vol. 6064, 332

\bibitem[\protect\citeauthoryear{{Heinis} et~al.}{{Heinis}
  et~al.}{2009}]{Heinis2009}
{Heinis}, S., et~al. 2009, \apj, 698, 1838

\bibitem[\protect\citeauthoryear{{Hogg}}{{Hogg}}{2001}]{Hogg2001}
{Hogg}, D.~W. 2001, \aj, 121, 1207

\bibitem[\protect\citeauthoryear{{Hogg} \& {Turner}}{{Hogg} \&
  {Turner}}{1998}]{Hogg1998}
{Hogg}, D.~W.,  \& {Turner}, E.~L. 1998, \pasp, 110, 727

\bibitem[\protect\citeauthoryear{{Hoversten} et~al.}{{Hoversten}
  et~al.}{2009}]{Hoversten2009}
{Hoversten}, E.~A., et~al. 2009, \apj, 705, 1462

\bibitem[\protect\citeauthoryear{{Ilbert} et~al.}{{Ilbert}
  et~al.}{2009}]{Ibert2009}
{Ilbert}, O., et~al. 2009, \apj, 690, 1236

\bibitem[\protect\citeauthoryear{{Jenkins} et~al.}{{Jenkins}
  et~al.}{2007}]{Jenkins2007}
{Jenkins}, L.~P., {Hornschemeier}, A.~E., {Mobasher}, B., {Alexander}, D.~M.,
  \& {Bauer}, F.~E. 2007, \apj, 666, 846

\bibitem[\protect\citeauthoryear{{Kennicutt}}{{Kennicutt}}{1998}]{KennicuttSF}
{Kennicutt}, R.~C. 1998, \araa, 36, 189

\bibitem[\protect\citeauthoryear{{Kinney} et~al.}{{Kinney}
  et~al.}{1996}]{Kinney1996}
{Kinney}, A.~L., {Calzetti}, D., {Bohlin}, R.~C., {McQuade}, K.,
  {Storchi-Bergmann}, T.,  \& {Schmitt}, H.~R. 1996, \apj, 467, 38

\bibitem[\protect\citeauthoryear{{Komiyama} et~al.}{{Komiyama}
  et~al.}{2002}]{Komiyama2002}
{Komiyama}, Y., et~al. 2002, \apjs, 138, 265

\bibitem[\protect\citeauthoryear{{Kron}}{{Kron}}{1980}]{Kron1980}
{Kron}, R.~G. 1980, \apjs, 43, 305

\bibitem[\protect\citeauthoryear{{{\L}okas} \& {Mamon}}{{{\L}okas} \&
  {Mamon}}{2003}]{Lokas2003}
{{\L}okas}, E.~L.,  \& {Mamon}, G.~A. 2003, \mnras, 343, 401

\bibitem[\protect\citeauthoryear{{Lupton} et~al.}{{Lupton}
  et~al.}{2001}]{Lupton2001}
{Lupton}, R., {Gunn}, J.~E., {Ivezi{\'c}}, Z., {Knapp}, G.~R.,  \& {Kent}, S.
  2001, in ASP-ADASS, ed. F.~R. {Harnden}, Jr., F.~A. {Primini}, \& H.~E.
  {Payne}, Vol. 238, 269

\bibitem[\protect\citeauthoryear{{Martin} et~al.}{{Martin}
  et~al.}{2005}]{Martin2005}
{Martin}, D.~C., et~al. 2005, \apjl, 619, L59

\bibitem[\protect\citeauthoryear{{Miller}, {Hornschemeier}, \&
  {Mobasher}}{{Miller} et~al.}{2009}]{Miller2009}
{Miller}, N.~A., {Hornschemeier}, A.~E.,  \& {Mobasher}, B. 2009, \aj, 137,
  4436

\bibitem[\protect\citeauthoryear{{Miller} \& {Owen}}{{Miller} \&
  {Owen}}{2002}]{Miller2002}
{Miller}, N.~A.,  \& {Owen}, F.~N. 2002, \aj, 124, 2453

\bibitem[\protect\citeauthoryear{{Milliard}, {Donas}, \& {Laget}}{{Milliard}
  et~al.}{1991}]{Milliard1991}
{Milliard}, B., {Donas}, J.,  \& {Laget}, M. 1991, Advances in Space Research,
  11, 135

\bibitem[\protect\citeauthoryear{{Milliard} et~al.}{{Milliard}
  et~al.}{1992}]{Milliard1992}
{Milliard}, B., {Donas}, J., {Laget}, M., {Armand}, C.,  \& {Vuillemin}, A.
  1992, \aap, 257, 24

\bibitem[\protect\citeauthoryear{{Milliard} et~al.}{{Milliard}
  et~al.}{2007}]{Milliard2007}
{Milliard}, B., et~al. 2007, \apjs, 173, 494

\bibitem[\protect\citeauthoryear{{Mobasher} et~al.}{{Mobasher}
  et~al.}{2001}]{Mobasher2001}
{Mobasher}, B., et~al. 2001, \apjs, 137, 279

\bibitem[\protect\citeauthoryear{{Morrissey} et~al.}{{Morrissey}
  et~al.}{2007}]{Morrissey2007}
{Morrissey}, P., et~al. 2007, \apjs, 173, 682

\bibitem[\protect\citeauthoryear{{Morrissey} et~al.}{{Morrissey}
  et~al.}{2005}]{Morrissey2005}
{Morrissey}, P., et~al. 2005, \apjl, 619, L7

\bibitem[\protect\citeauthoryear{{Nagashima} et~al.}{{Nagashima}
  et~al.}{2002}]{Nagashima2002}
{Nagashima}, M., {Yoshii}, Y., {Totani}, T.,  \& {Gouda}, N. 2002, \apj, 578,
  675

\bibitem[\protect\citeauthoryear{{Neumann} et~al.}{{Neumann}
  et~al.}{2001}]{Neumann2001}
{Neumann}, D.~M., et~al. 2001, \aap, 365, L74

\bibitem[\protect\citeauthoryear{{O'Connell}}{{O'Connell}}{1999}]{Oconnell1999}
{O'Connell}, R.~W. 1999, \araa, 37, 603

\bibitem[\protect\citeauthoryear{{Park} \& {Lee}}{{Park} \&
  {Lee}}{1997}]{Park1997}
{Park}, J.-H.,  \& {Lee}, Y.-W. 1997, \apj, 476, 28

\bibitem[\protect\citeauthoryear{{Pimbblet}, {Drinkwater}, \&
  {Hawkrigg}}{{Pimbblet} et~al.}{2004}]{Pimbblet2004}
{Pimbblet}, K.~A., {Drinkwater}, M.~J.,  \& {Hawkrigg}, M.~C. 2004, \mnras,
  354, L61

\bibitem[\protect\citeauthoryear{{Poggianti} et~al.}{{Poggianti}
  et~al.}{2004}]{Poggianti2004}
{Poggianti}, B.~M., {Bridges}, T.~J., {Komiyama}, Y., {Yagi}, M., {Carter}, D.,
  {Mobasher}, B., {Okamura}, S.,  \& {Kashikawa}, N. 2004, \apj, 601, 197

\bibitem[\protect\citeauthoryear{{Rawle} et~al.}{{Rawle}
  et~al.}{2008}]{Rawle2008}
{Rawle}, T.~D., {Smith}, R.~J., {Lucey}, J.~R., {Hudson}, M.~J.,  \& {Wegner},
  G.~A. 2008, \mnras, 385, 2097

\bibitem[\protect\citeauthoryear{{Salim} et~al.}{{Salim}
  et~al.}{2009}]{Salim2009}
{Salim}, S., et~al. 2009, \apj, 700, 161

\bibitem[\protect\citeauthoryear{{Schlegel}, {Finkbeiner}, \&
  {Davis}}{{Schlegel} et~al.}{1998}]{Schlegel1998}
{Schlegel}, D.~J., {Finkbeiner}, D.~P.,  \& {Davis}, M. 1998, \apj, 500, 525

\bibitem[\protect\citeauthoryear{{Smith} et~al.}{{Smith}
  et~al.}{2009}]{Smith2009}
{Smith}, R.~J., {Lucey}, J.~R., {Hudson}, M.~J., {Allanson}, S.~P., {Bridges},
  T.~J., {Hornschemeier}, A.~E., {Marzke}, R.~O.,  \& {Miller}, N.~A. 2009,
  \mnras, 392, 1265

\bibitem[\protect\citeauthoryear{{Stoughton} et~al.}{{Stoughton}
  et~al.}{2002}]{Stoughton2002}
{Stoughton}, C., et~al. 2002, \aj, 123, 485

\bibitem[\protect\citeauthoryear{{Teplitz} et~al.}{{Teplitz}
  et~al.}{2006}]{Teplitz2006}
{Teplitz}, H.~I., et~al. 2006, \aj, 132, 853

\bibitem[\protect\citeauthoryear{{White}, {Briel}, \& {Henry}}{{White}
  et~al.}{1993}]{White1993}
{White}, S.~D.~M., {Briel}, U.~G.,  \& {Henry}, J.~P. 1993, \mnras, 261, L8

\bibitem[\protect\citeauthoryear{{Wyder} et~al.}{{Wyder}
  et~al.}{2007}]{Wyder2007}
{Wyder}, T.~K., et~al. 2007, \apjs, 173, 293

\bibitem[\protect\citeauthoryear{{Wyder} et~al.}{{Wyder}
  et~al.}{2005}]{Wyder2005}
{Wyder}, T.~K., et~al. 2005, \apjl, 619, L15

\bibitem[\protect\citeauthoryear{{Xu} et~al.}{{Xu} et~al.}{2005}]{Xu2005}
{Xu}, C.~K., et~al. 2005, \apjl, 619, L11

\bibitem[\protect\citeauthoryear{{York} et~al.}{{York} et~al.}{2000}]{York2000}
{York}, D.~G., et~al. 2000, \aj, 120, 1579

\bibitem[\protect\citeauthoryear{{Zamojski} et~al.}{{Zamojski}
  et~al.}{2007}]{Zamojski2007}
{Zamojski}, M.~A., et~al. 2007, \apjs, 172, 468

\end{thebibliography}
\clearpage

\begin{figure}[t!]
\centerline{\scalebox{0.537}{\rotatebox{0.0}{\includegraphics*[37,146][575,686]{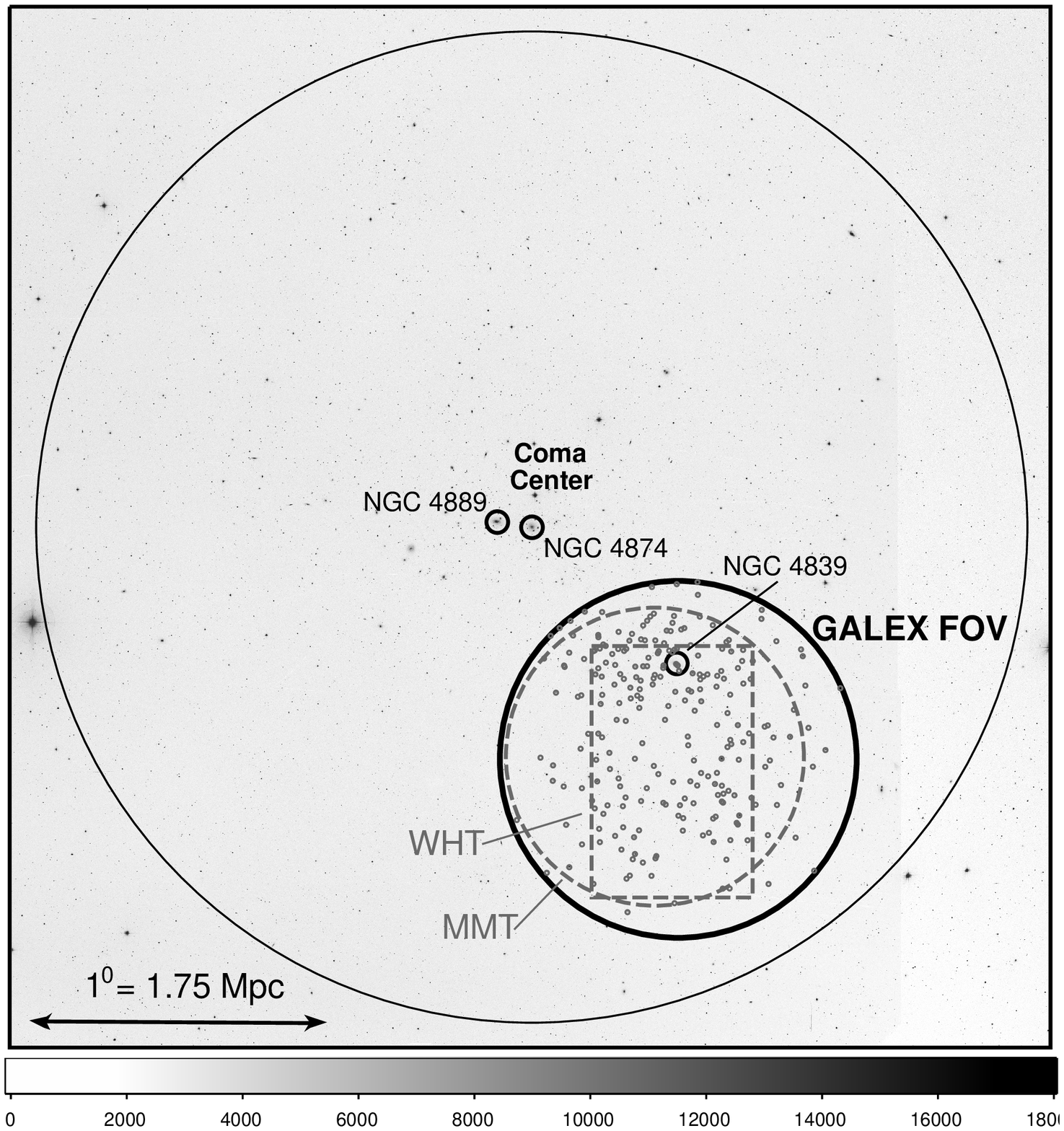}}}}
\centerline{\scalebox{0.80}{\rotatebox{-90.0}{\includegraphics*[120,240][476,602]{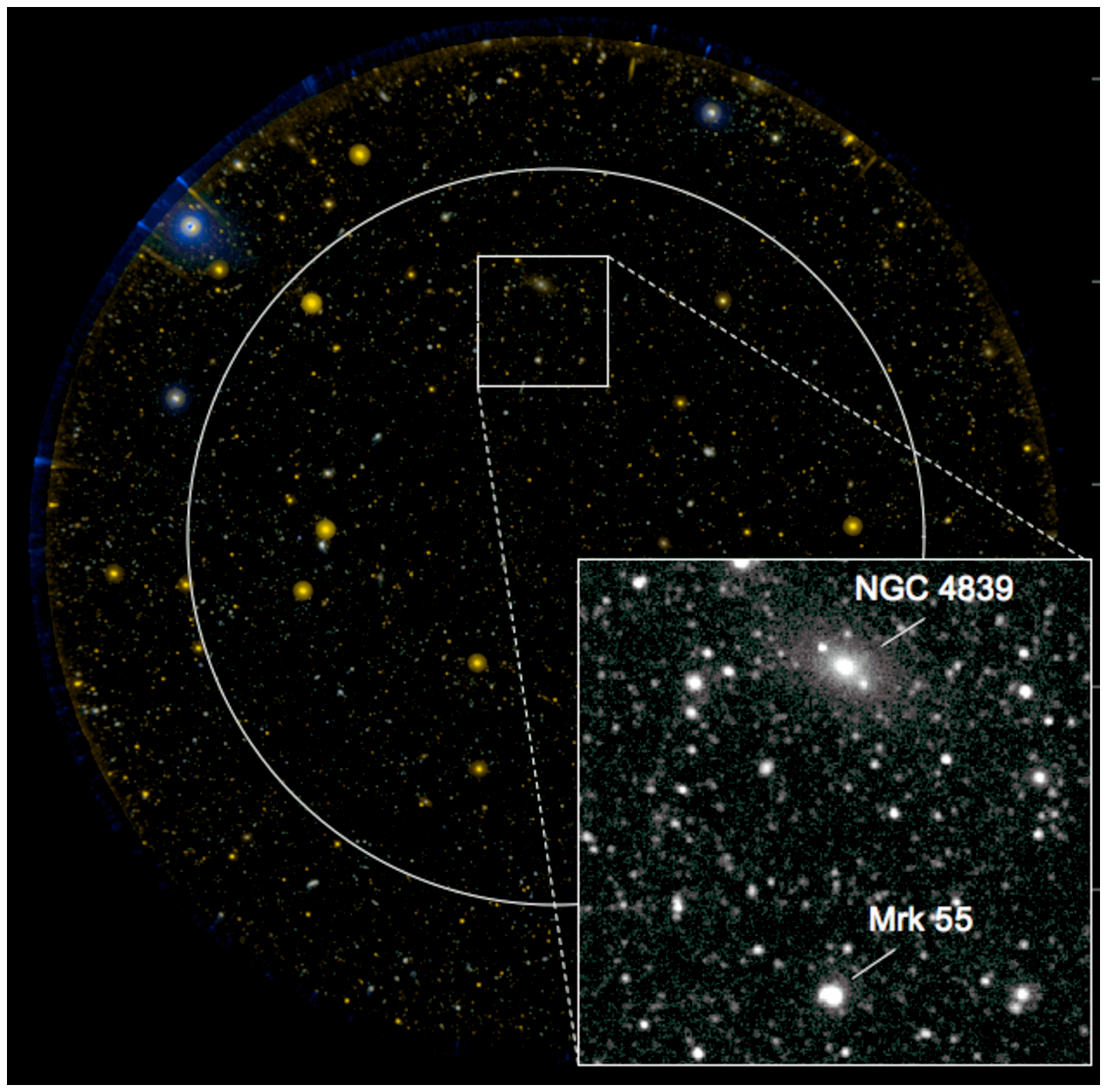}}}}
\caption{\label{coma_field}{\it Top}: Digitized Sky Survey image of the Coma cluster that extends slightly beyond the cluster virial radius \citep[the virial radius is indicated by a large circle, r=2.9 Mpc;][]{Lokas2003}.
The thick black circle (radius=0.6$^{\circ}$) shows the location of the {\it GALEX} field studied here.
Small gray dots show the location of 244 spectroscopically-confirmed Coma member galaxies with {\it GALEX}/SDSS photometry.
The gray dashed circle/rectangle are the spectroscopic footprints for deep redshift surveys performed with MMT-Hectospec (R.~Marzke et al.~2010, {\it in prep}) and from the \cite{Mobasher2001} survey, respectively.
We also mark the locations of the three largest galaxies in the Coma cluster (NGC 4889, 4874, 4839; small black circles).
{\it Bottom}: A two-color {\it GALEX} image (FUV=blue \& NUV=red) of the Coma cluster.
The inset (white square) shows a $9\arcmin$x$9\arcmin$ subregion of the pipeline-reduced
NUV image, which includes a bright Coma member galaxy (NGC 4839) and a LINER \citep[Mrk 55;][]{Miller2002}.
The depth of our 26 ks {\it GALEX} observation is demonstrated in the inset image.}
\end{figure}
\clearpage

\begin{figure}[t!]
\centerline{
\scalebox{0.3}{\rotatebox{0.}{\includegraphics*[10,100][602,692]{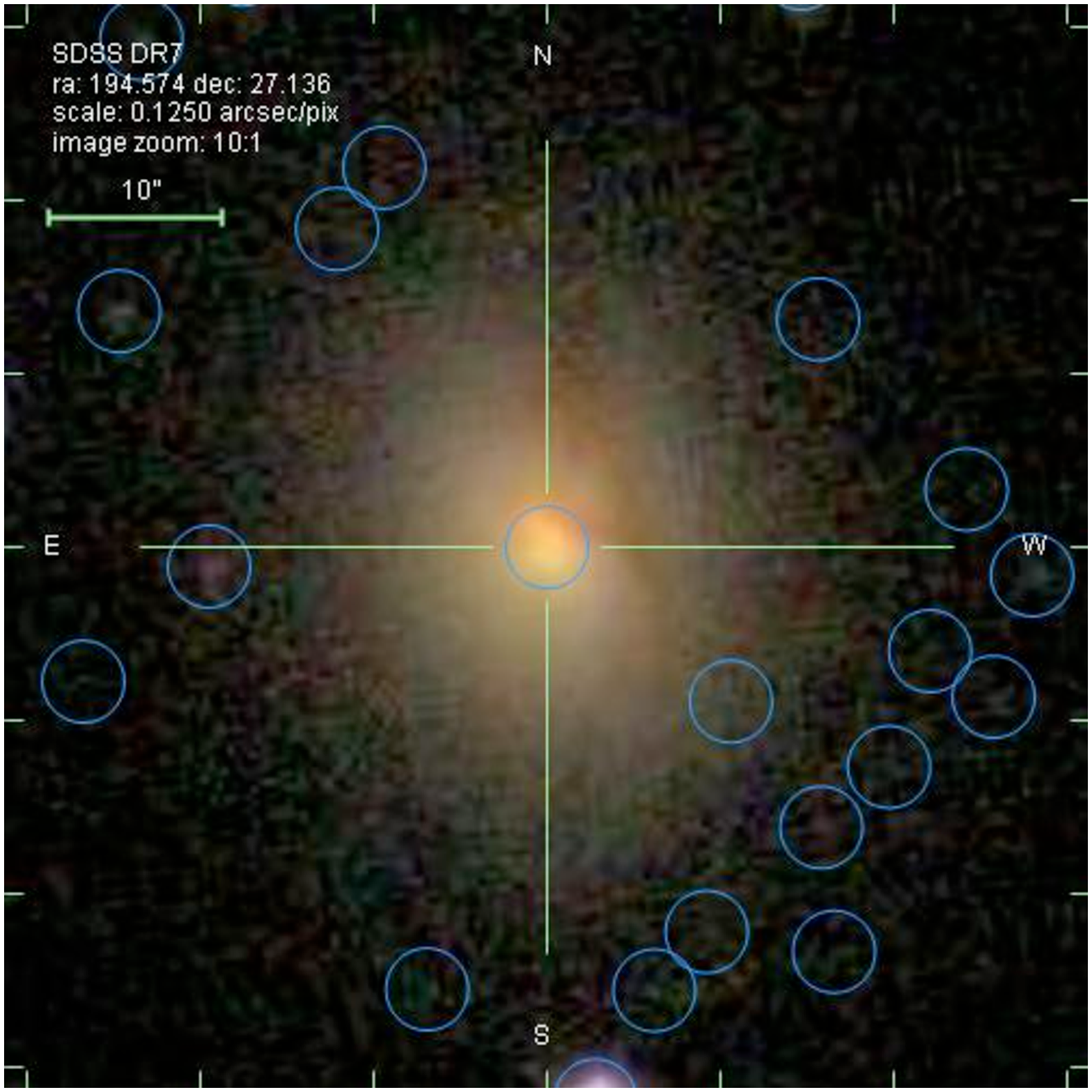}}}
\scalebox{0.3}{\rotatebox{0.}{\includegraphics*[10,100][602,692]{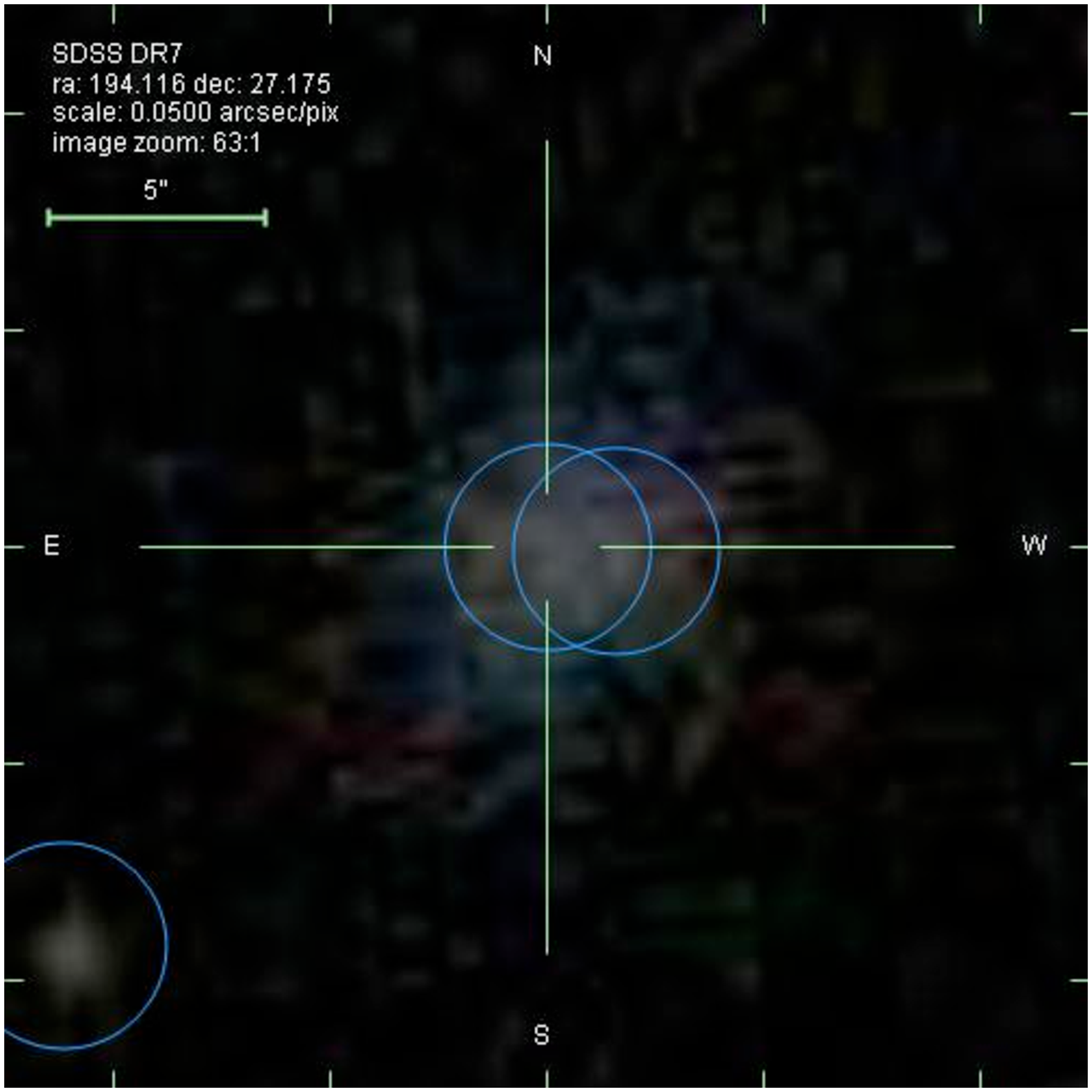}}}
\scalebox{0.3}{\rotatebox{0.}{\includegraphics*[10,100][602,692]{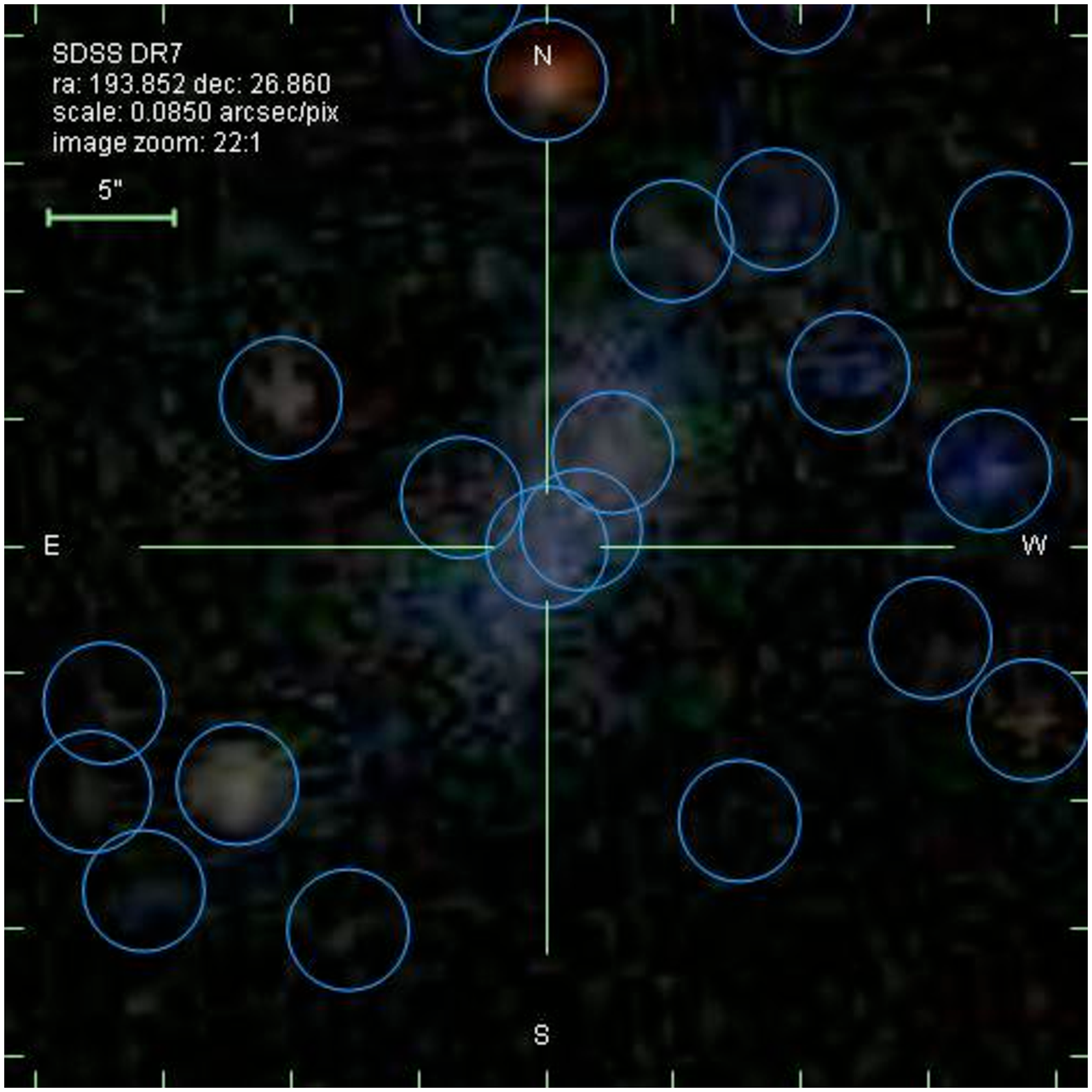}}}}
\caption{\label{sdss_blends}SDSS co-added ($ugriz$) cutout images for three galaxies in the Coma cluster that are examples of `shredding',
i.e.~a single galaxy is separated into two or more objects. The blue circles indicate individual sources (galaxies or stars) in the SDSS optical catalog.
Shredded sources in the Coma field are often associated with cluster members, 
particularly in the halos of bright early-type galaxies where multiple shreds are found ({\it left}; lower-right section of image).
Shredding is also frequently associated with low surface brightness (LSB) galaxies ({\it center} and {\it right}).}
\end{figure}
\clearpage

\begin{figure}[t!]
\centering
\includegraphics[scale=0.75]{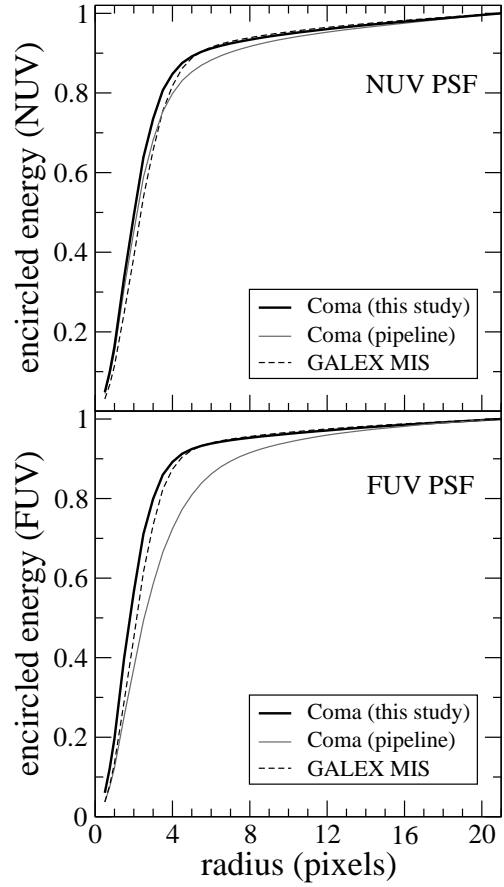}
\caption{\label{galex_psf} Encircled energy curves of the {\it GALEX} PSF for the NUV and FUV filters.
In each panel we show (a) the PSF that we measured using stars located in our {\it GALEX} image (black solid line),
(b) the PSF automatically generated by the {\it GALEX} pipeline for our image (gray solid line), and 
(c) the average PSF measured by the {\it GALEX} team for the MIS survey (black dashed).
The pipeline-generated PSF has a relatively broad profile due to the contamination of its star sample by nearby objects.}
\end{figure}
\clearpage

\begin{figure}[t!]
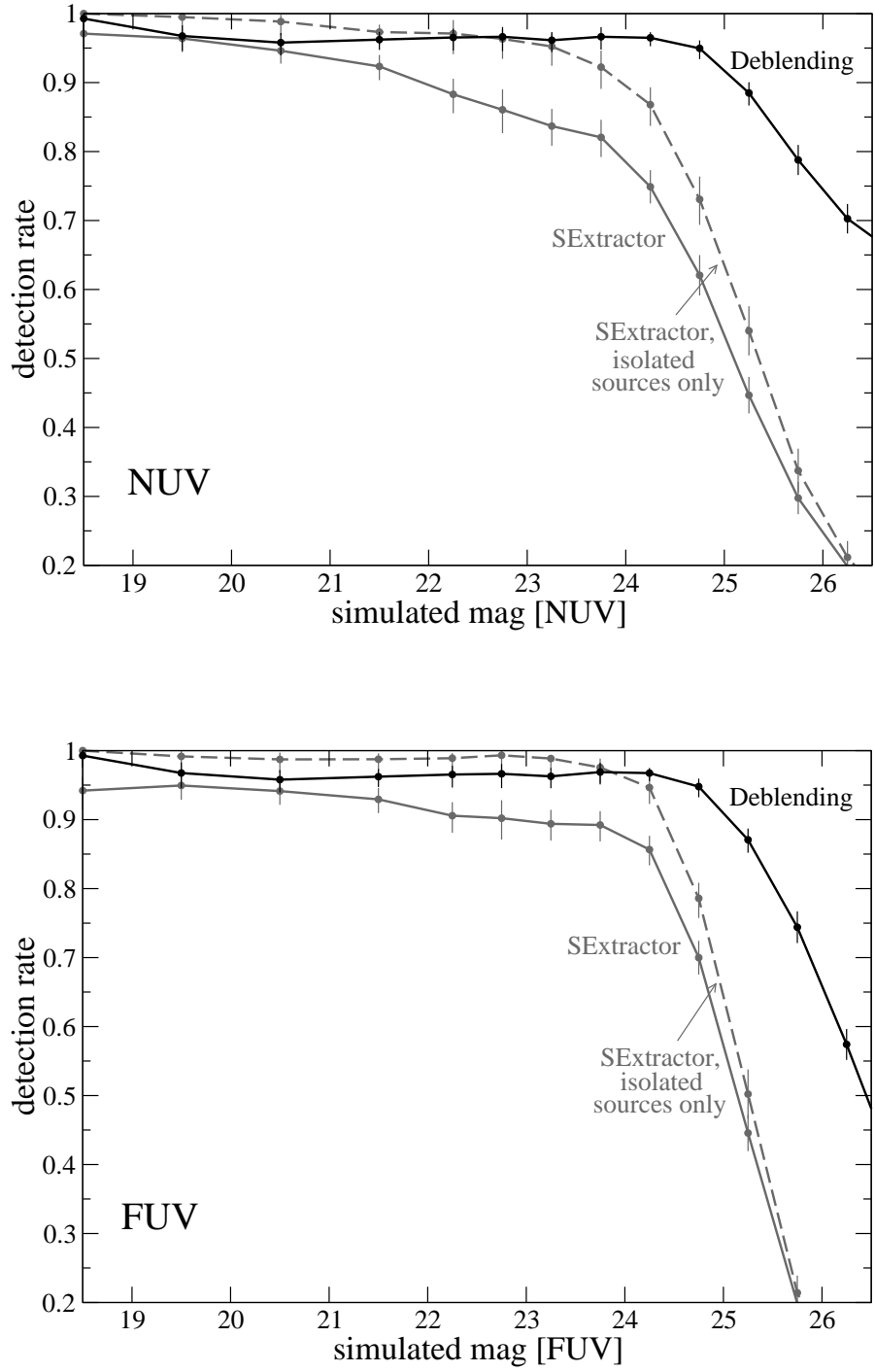

\centering
\includegraphics[scale=0.75]{f4a.eps}
\\[40pt]
\includegraphics[scale=0.75]{f4b.eps}
\caption{\label{deteff} Detection rates for $\sim$4000 artificial point sources that were inserted in the {\it GALEX} pipeline NUV (top) and FUV (bottom) images.
Source detection was performed using a Bayesian deblending algorithm (solid black lines) and SExtractor (solid gray lines) with parameters identical to the {\it GALEX} pipeline.
The SExtractor detection efficiency is measured after matching the SExtractor output catalog to the SDSS+artificial source catalog using a 4$\arcsec$ search radius.
The dashed gray lines show the SExtractor detection rates for the subset of simulated sources that are {\it isolated} (i.e.~no SDSS source is located within 10$\arcsec$);
isolated sources are free from object blends, which results in a $\sim$10\% higher detection rate at $NUV$=23 and $FUV$=24 as compared to the full simulated sample.
The isolated detection rates for the Bayesian deblending algorithm (not shown) are nearly identical to the full sample.}
\end{figure}
\clearpage

\begin{figure}[t!]
\centering
\includegraphics[scale=0.75]{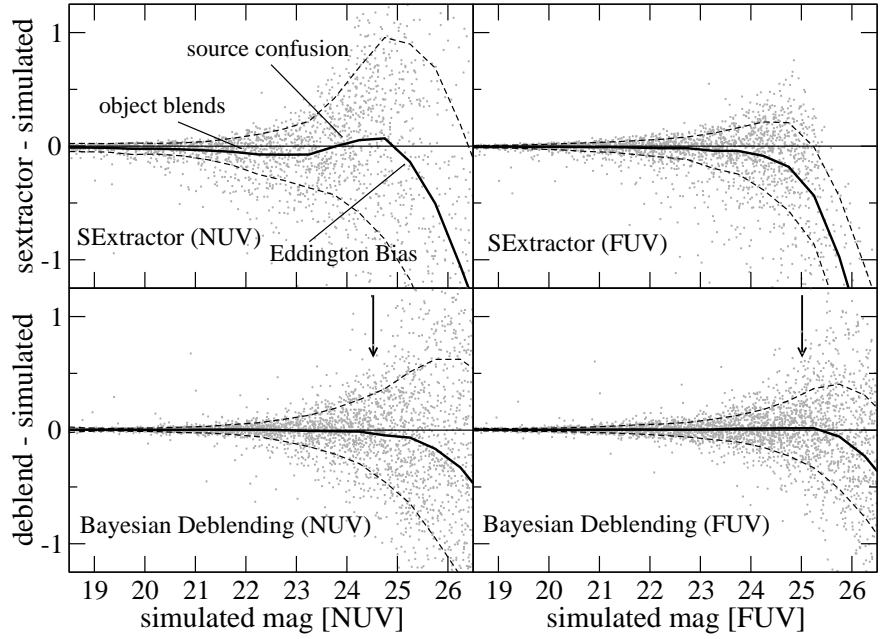}
\caption{\label{magdiff} Comparison of artificial point source magnitudes to the measurements performed using a Bayesian deblending technique and also using 
SExtactor (with a configuration that is identical to the {\it GALEX} pipeline).
The 3$\sigma$ clipped average $\pm$ standard deviation are shown in each panel (solid and dashed lines, respectively).
We have applied a small aperture correction to all SExtractor magnitudes in Figure \ref{magdiff} ($\Delta$$NUV$=0.1 and $\Delta$$FUV$=0.07).
We identify three regions in the SExtractor NUV comparison that are affected by other systematic errors (object blends, source confusion, and the Eddington Bias).
Arrows indicate our chosen photometry limits for the Bayesian deblending catalog ($NUV$=24.5 and $FUV$=25.0).}
\end{figure}
\clearpage

\begin{figure}[t!]
\centering
\includegraphics[scale=0.75]{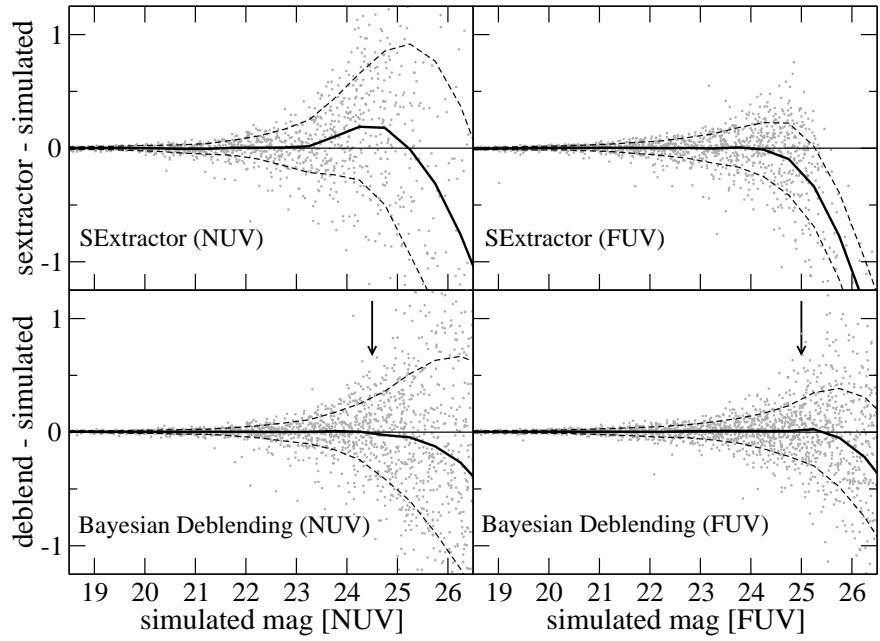}
\caption{\label{magdiff_ISO} Same magnitude comparison as shown in Figure \ref{magdiff} but plotted for isolated sources alone (i.e.~we discarded simulated objects located within 10$\arcsec$ of any SDSS source).
This removes systematic offsets due to object blends. Arrows indicate our chosen photometry limits for the Bayesian deblending catalog ($NUV$=24.5 and $FUV$=25.0).}
\end{figure}
\clearpage

\begin{figure}[t!]
\centering
\includegraphics[scale=0.75]{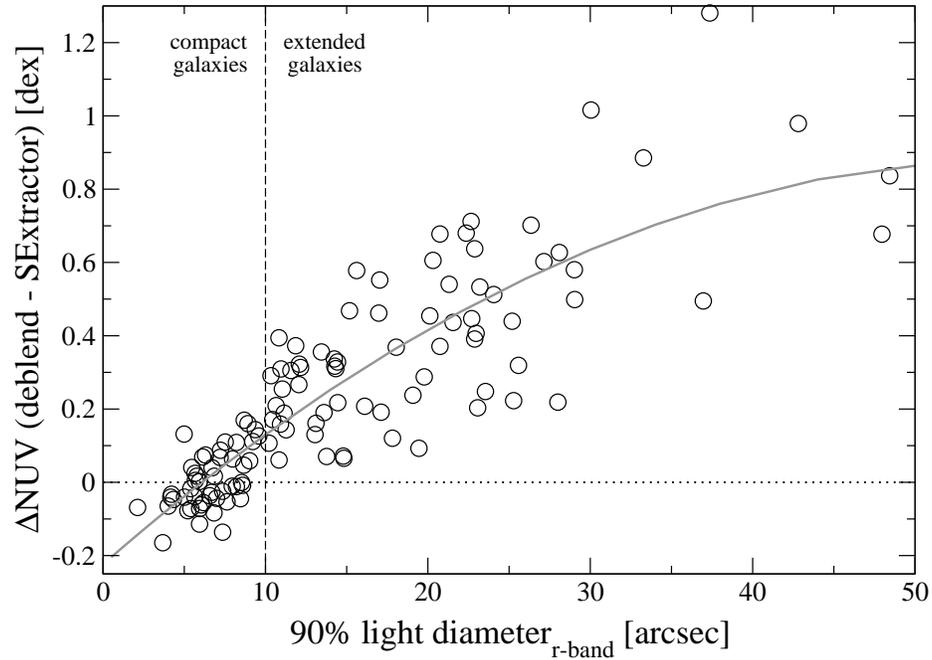}
\caption{\label{magdiff_ext} Photometry comparison for galaxies in both our Bayesian deblending and {\it GALEX} pipeline catalogs, versus the size of the galaxy
(taken as the optical 90\% light diameter in the SDSS $r$-band).
It is expected that since Bayesian deblending assumes a PSF-like light profile, it should underestimate the flux for extended galaxies that have relatively broad light profiles.
We restricted this comparison to bright galaxies ($NUV$$<$21) and isolated galaxies (i.e.~no additional SDSS sources are located within 10$\arcsec$) in order to limit the effects of measurement error.
The long dashed line shows our chosen size limit to separate compact and extended galaxies for which we adopt photometry from the Bayesian deblending catalog and the {\it GALEX} pipeline catalog, respectively.
The magnitude offsets have a dispersion of 0.07 and 0.18 dex for galaxies smaller/larger than 10$\arcsec$, respectively, based on 2nd-order polynomial fit to the data (solid gray line).
The dotted line corresponds to a magnitude offset equal to zero.}
\end{figure}
\clearpage

\begin{figure}[t!]
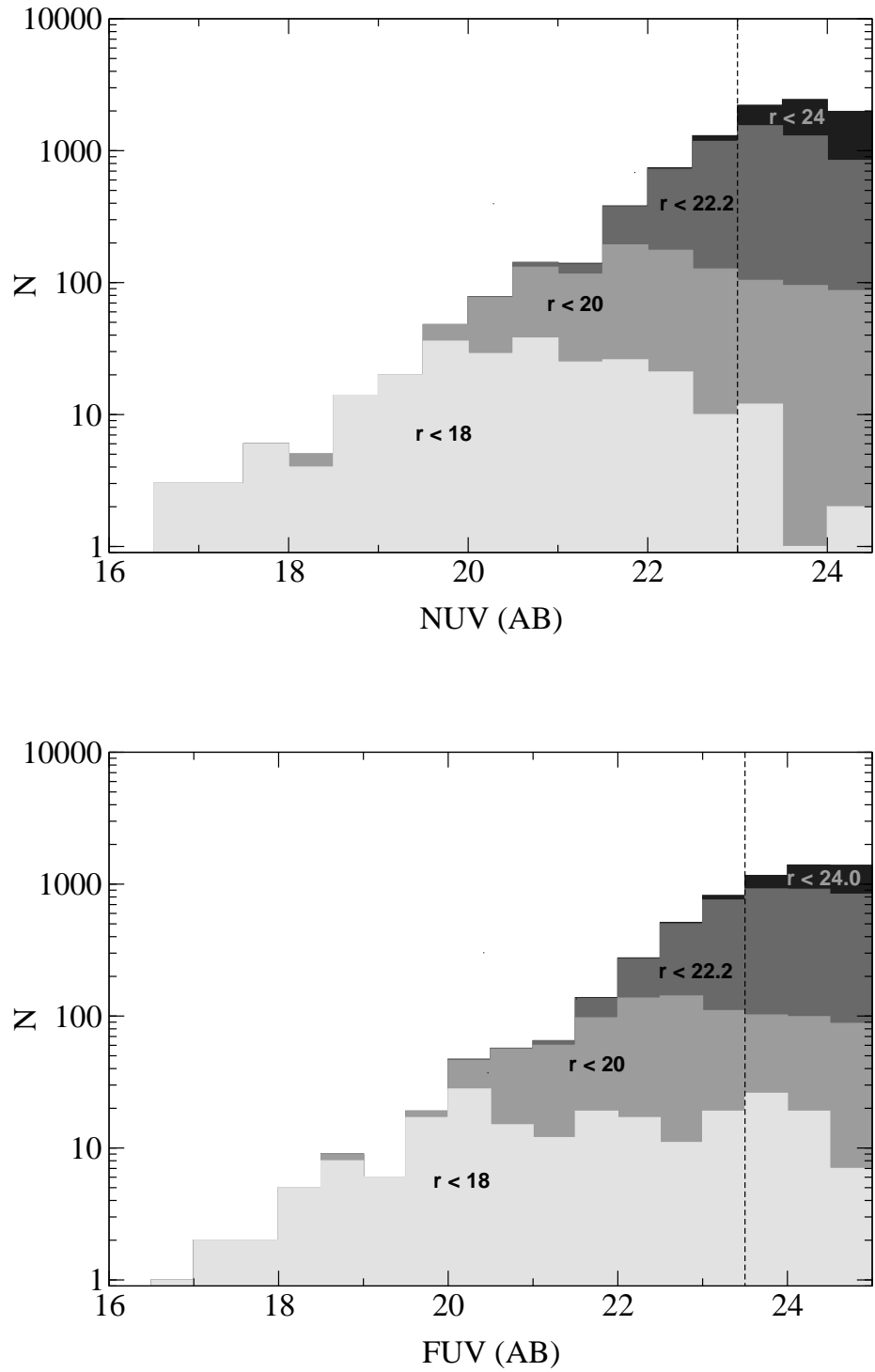

\centering
\includegraphics[scale=0.75]{f8a.eps}
\\[40pt]
\includegraphics[scale=0.75]{f8b.eps}
\caption{\label{uvhisto} NUV (top) and FUV (bottom) magnitude distribution for all sources in the {\it GALEX}/SDSS catalog.
We have separated sources by the SDSS $r$-band magnitude (shaded).
The dashed vertical lines indicate the UV completeness limit adopted for the combined {\it GALEX}/SDSS catalog.}
\end{figure}
\clearpage

\begin{figure}[t!]
\centering
\includegraphics[scale=0.75]{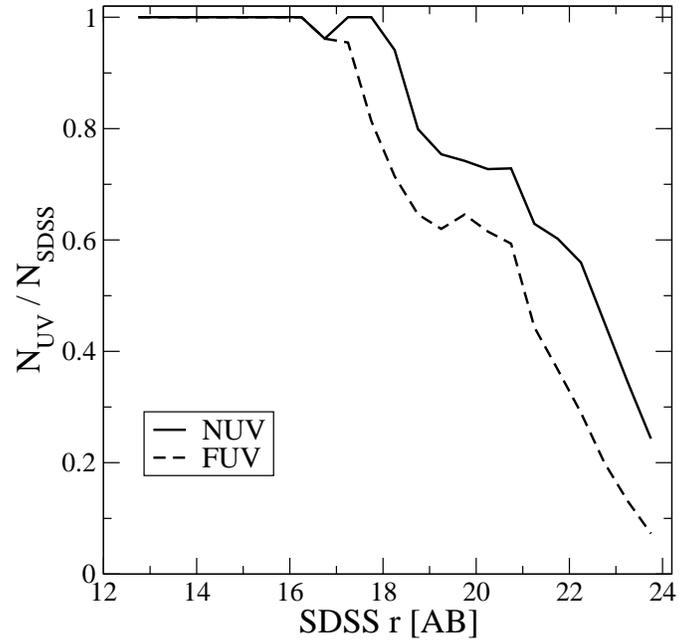}
\caption{\label{uvfrac}
Fraction of sources in the SDSS DR6 photometric catalog that are detected in our {\it GALEX} image at magnitudes brighter than our 
UV photometry limits ($NUV$=24.5 and $FUV$=25.0).
The solid and dashed lines show the fraction of SDSS sources with NUV and FUV detections, respectively.
SDSS sources (16,759 total) were selected from the inner 0.5$^{\circ}$ {\it GALEX} FOV.
A single SDSS galaxy at $r$$\sim$17 was not detected in the {\it GALEX} NUV image due to a nearby UV-bright star.
The cusp in the UV-detection fraction across $r$=19-21 corresponds to the magnitude where blue star forming galaxies
dominate the galaxy population, as opposed to quiescent red galaxies which are the majority galaxy-type at brighter magnitudes.}
\end{figure}
\clearpage

\begin{figure}[t!]
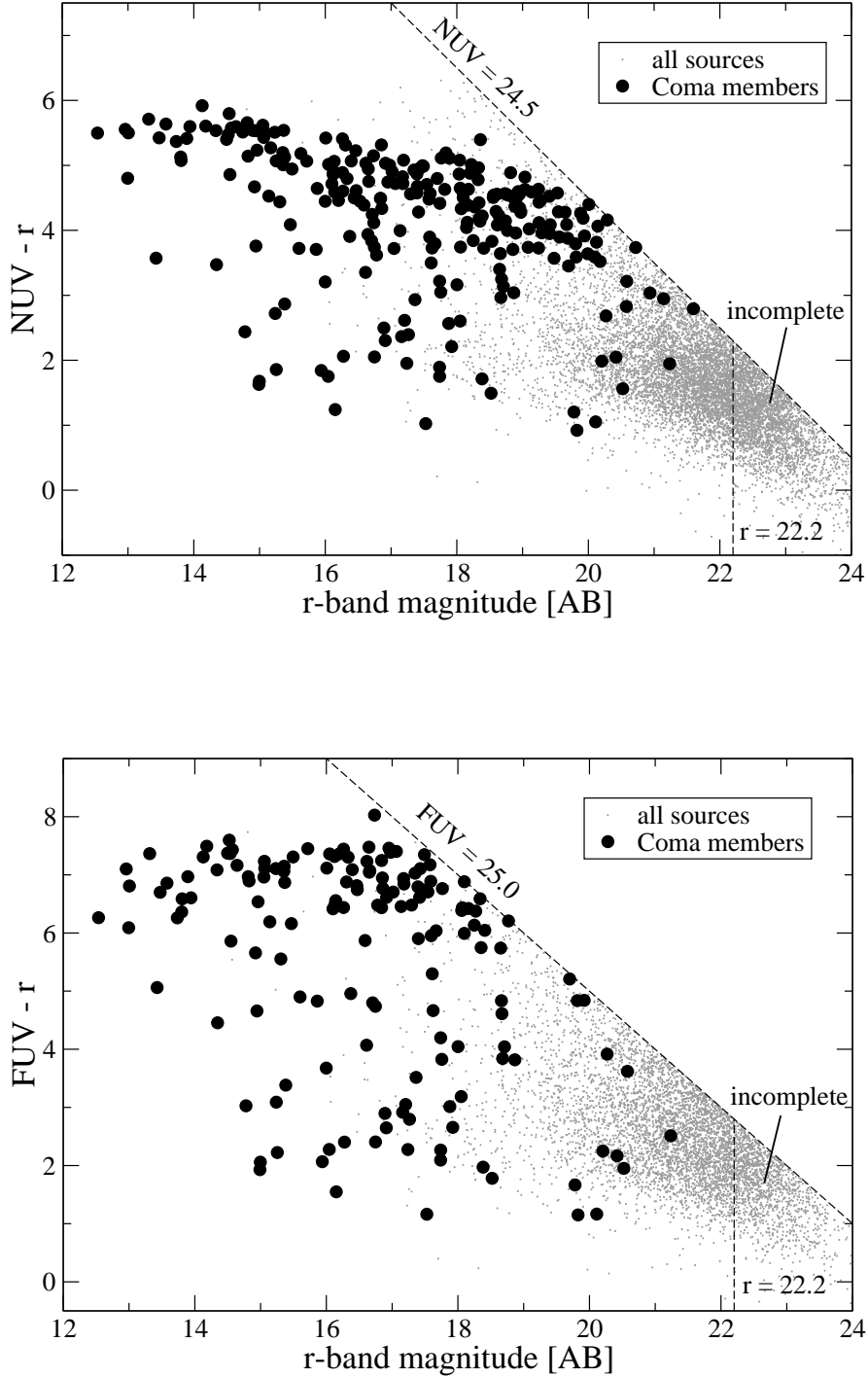

\centering
\includegraphics[scale=0.75]{f10a.eps}
\\[50pt]
\includegraphics[scale=0.75]{f10b.eps}
\caption{\label{colormag} Color-magnitude diagram (CMD) for all NUV-detected sources ({\it top}) and FUV-detected sources ({\it bottom}) in our {\it GALEX}/SDSS catalog.
Small gray dots show all sources in our {\it GALEX}/SDSS catalog, and large filled circles 
indicate spectroscopically-confirmed Coma members (4000 km s$^{-1}$$\leq$$cz$$\leq$10,000 km s$^{-1}$).
The long dashed lines indicate the reliable photometry limits for the {\it GALEX} image ($NUV$=24.5; $FUV$=25.0), and the 95\% completeness limit of the SDSS catalog \citep[$r$=22.2;][]{Adelman2008}.}
\end{figure}
\clearpage

\begin{figure}[t!]
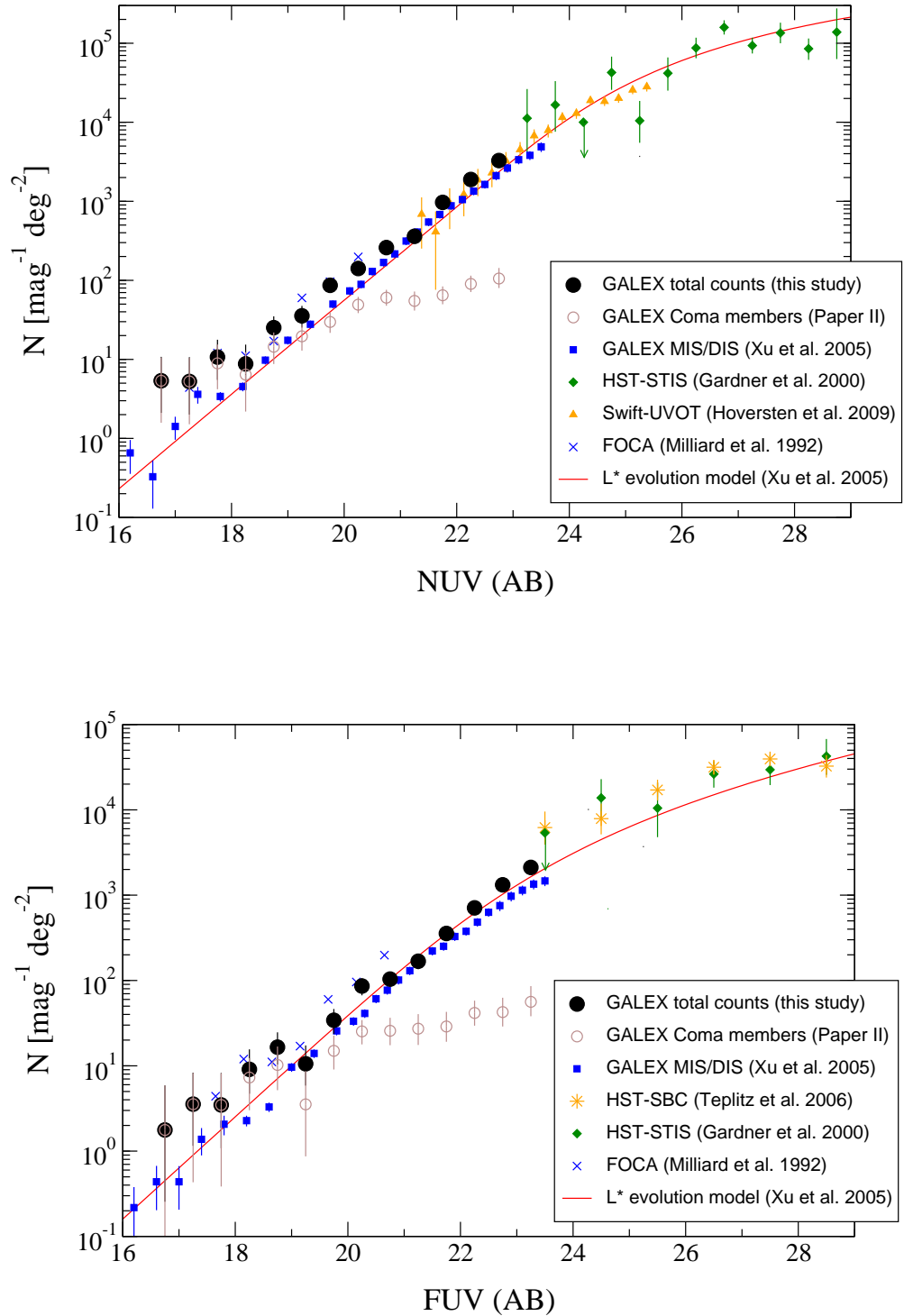

\centering
\includegraphics[angle=0.,scale=0.75]{f11a.eps}
\\[45pt]
\includegraphics[angle=0.,scale=0.75]{f11b.eps}
\caption{\label{uvdiff} NUV (top panel) and FUV (bottom panel) differential galaxy counts for the total population (cluster+background; bold filled circles).
Coma member galaxies (open brown circles) are only a small fraction of the total galaxy counts at magnitudes fainter than $NUV$$\approx$21 and $FUV$$\approx$21.
Other symbols show bright galaxy counts observed with FOCA at 2000 \AA~\citep[X signs;][]{Milliard1992}, a previous {\it GALEX} NUV and FUV study \citep[blue squares;][]{Xu2005},
a UVOT NUV study in the uvm2 band \citep[orange triangles;][]{Hoversten2009},
a deep HST-STIS NUV and FUV study \citep[green diamonds;][]{Gardner2000}, and a deep HST-SBC FUV study \citep[orange stars;][]{Teplitz2006}.
The red line shows the luminosity evolution model presented in \cite{Xu2005}.}
\end{figure}
\clearpage

\begin{figure}[t!]
\centering
\includegraphics[angle=0.,scale=0.75]{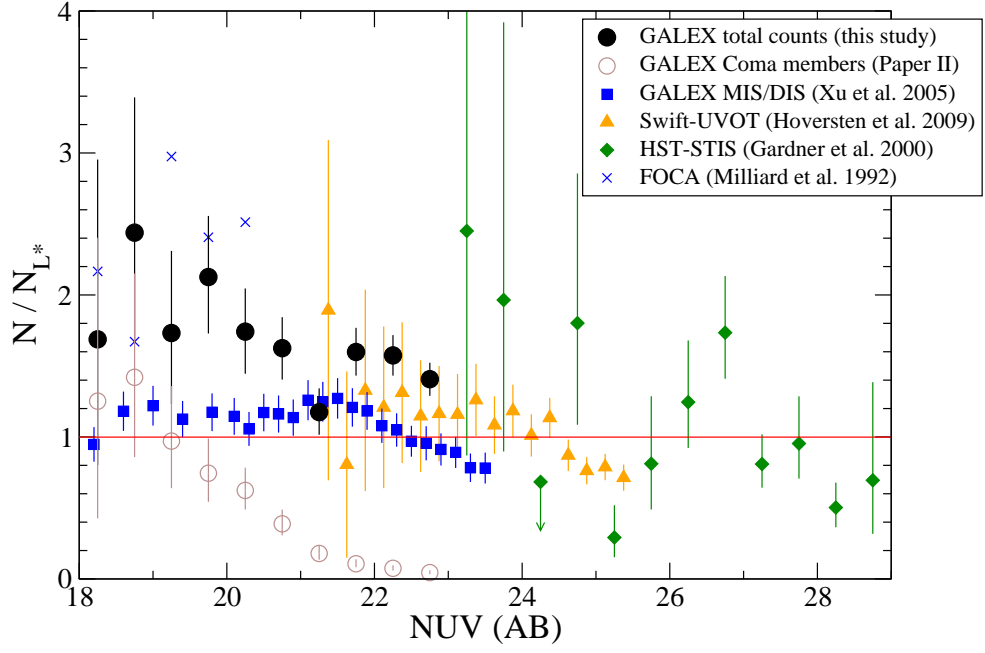}
\\[45pt]
\includegraphics[angle=0.,scale=0.75]{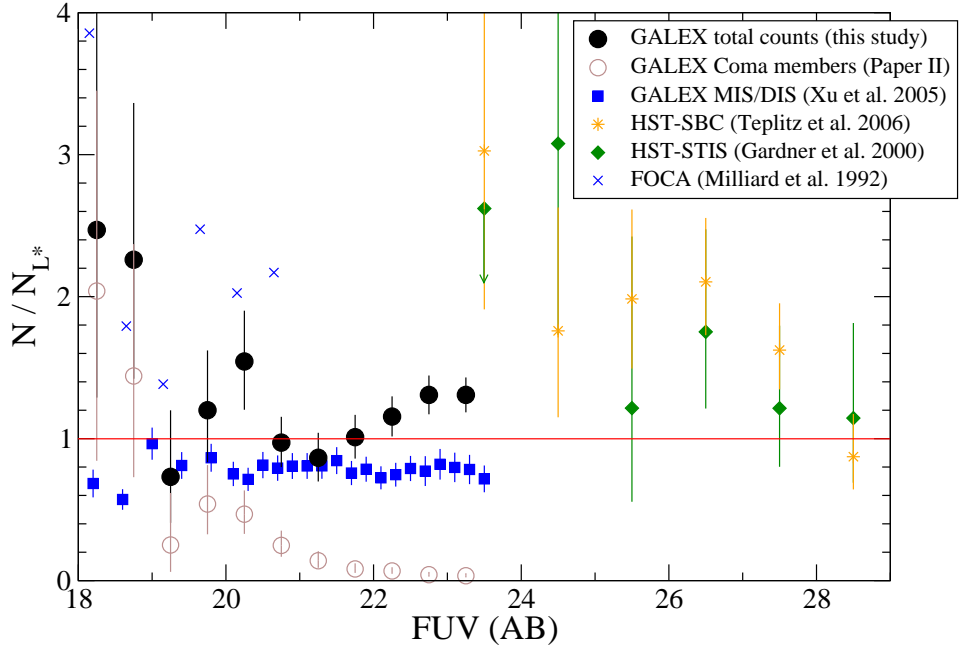}
\caption{\label{uvdiff_baseline}Same as Figure \ref{uvdiff} with differential galaxy counts normalized to the luminosity evolution model presented in \cite{Xu2005}.}
\end{figure}
\clearpage

\begin{figure}[t!]
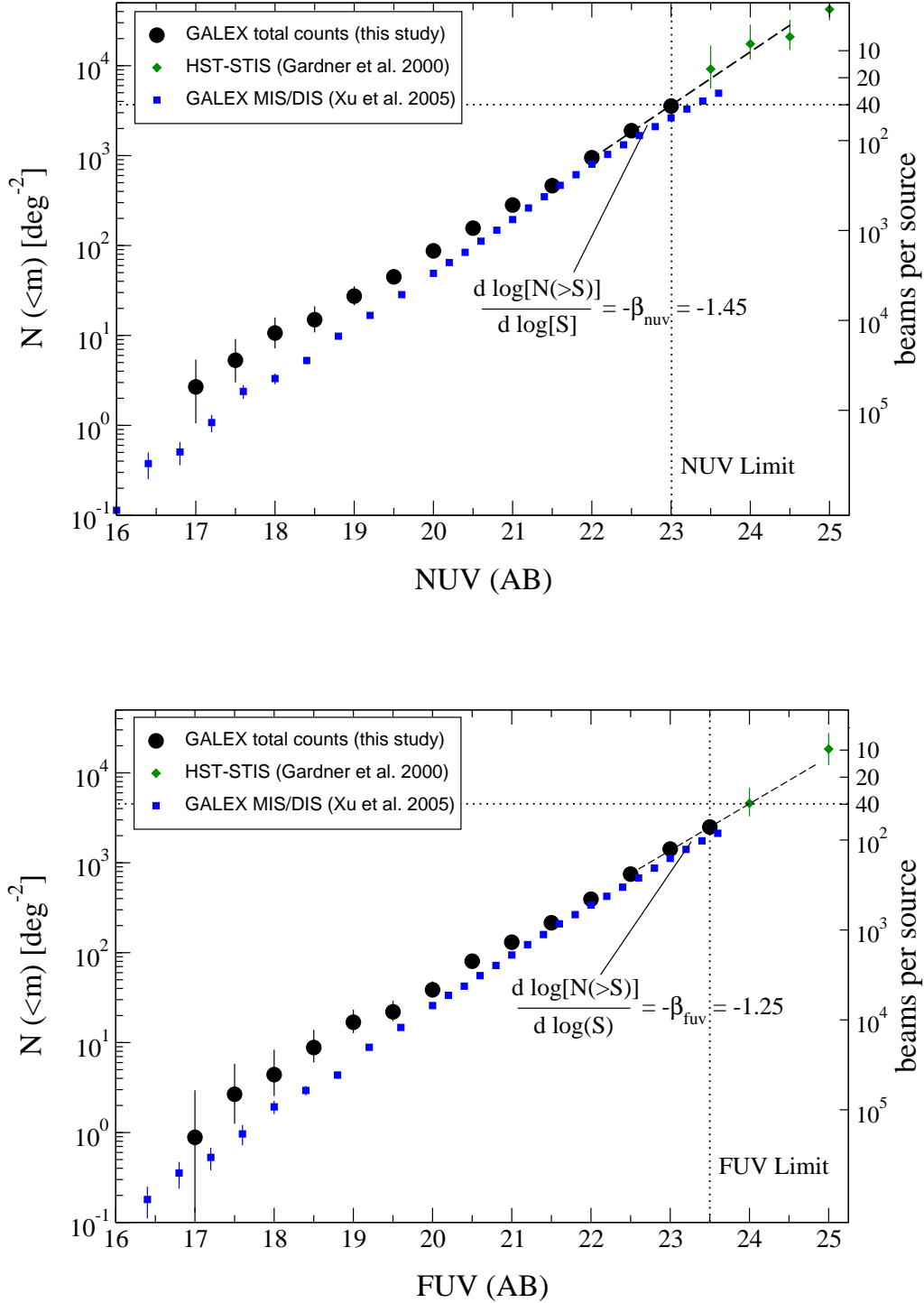

\centering
\includegraphics[angle=0.,scale=0.75]{f13a.eps}
\\[45pt]
\includegraphics[angle=0.,scale=0.75]{f13b.eps}
\caption{\label{uvdiff_cumul} NUV (top panel) and FUV (bottom panel) cumulative galaxy counts for our {\it GALEX}/SDSS catalog (bold filled circles).
The cumulative galaxy counts are the integral of the differential galaxy counts shown in Figure \ref{uvdiff}; symbols are the same as in Figure \ref{uvdiff}.
The vertical dotted lines are the chosen UV completeness limits of our catalog ($NUV$=23 and $FUV$=23.5).
The HST-STIS counts \citep{Gardner2000} were added to our cumulative galaxy counts at NUV (FUV)=23.0.
The dashed lines are linear fits to the log number counts across the three faintest magnitude bins.
The measured slopes are $\beta$=1.25 and 1.45 for the FUV and NUV bands, respectively (converted to a slope measured in units of flux).
The right vertical axis shows the number of beams-per-source.
We estimate that the 40 beams-per-source confusion limits for the {\it GALEX} pipeline catalogs are $NUV$=23.0 and $FUV$=24.0.}
\end{figure}
\clearpage

\begin{deluxetable}{cccccccccccc}
\tablecolumns{12}
\tabletypesize{\scriptsize}
\tablecaption{\label{Catalog} Properties of the {\it GALEX}/SDSS Catalog in Coma}
\tablewidth{0pt}
\tablenum{1}
\tablehead{
\colhead{R.A.} & \colhead{Dec.} & \colhead{S/G Type} & \colhead{$u$} & \colhead{$g$} & \colhead{$r$} & \colhead{$i$} & \colhead{$z$} & \colhead{$NUV$} & \colhead{$\sigma$$_{NUV}$} & \colhead{$FUV$} & \colhead{$\sigma$$_{FUV}$} \\
\colhead{(1)} & \colhead{(2)} & \colhead{(3)} & \colhead{(4)} & \colhead{(5)} & \colhead{(6)} & \colhead{(7)} & \colhead{(8)} & \colhead{(9)} & \colhead{(10)} & \colhead{(11)} & \colhead{(12)}} \\
\startdata
194.57009 & 27.72968  & 3 & 18.00 & 16.67 & 16.07 & 15.75 & 15.54 & 18.6 & 0.05 & 19.2 & 0.05 \\
194.69751 & 27.67473  & 3 & 16.56 & 14.62 & 13.81 & 13.44 & 13.25 & 18.9 & 0.05 & 20.4 & 0.06 \\
194.63167 & 27.67351  & 3 & 18.06 & 16.39 & 15.63 & 15.28 & 15.01 & 20.8 & 0.05 & 22.7 & 0.10 \\
194.73922 & 27.65435  & 3 & 20.18 & 19.42 & 18.59 & 18.31 & 18.07 & 20.8 & 0.05 & 21.6 & 0.06 \\
194.75167 & 27.64399  & 3 & 19.64 & 18.77 & 18.52 & 18.39 & 18.29 & 20.0 & 0.05 & 20.3 & 0.05 \\
\nodata	  &	      &   &       &       &       &       &       &      &     &      &		\\
\enddata
\tablecomments{See the electronic version of this paper for the complete catalog.}
\end{deluxetable}

\begin{deluxetable}{cllllc}
\tablecolumns{6}
\tabletypesize{\scriptsize}
\tablecaption{UV Galaxy Number Counts in the Coma Field \label{NumberCounts}}
\tablewidth{0pt}
\tablenum{2}
\tablehead{
\colhead{AB mag} & \colhead{$dN/dm$} & \colhead{$\sigma_{high}$} & \colhead{$\sigma_{low}$} & \colhead{$N_{raw}$} & \colhead{area} \\
\colhead{(1)} & \colhead{(2)} & \colhead{(3)} & \colhead{(4)} & \colhead{(5)} & \colhead{(6)} } \\
\startdata
\cutinhead{NUV}
16.75 & 5.4    & 5.4   & 3.3	& 3	& 1.13	\\
17.25 & 5.2    & 5.4   & 3.2	& 3   	& 1.13	\\
17.75 & 10.7    & 7.0   & 5.2	& 6	& 1.13	\\
18.25 & 8.6   & 6.5   & 4.5	& 5	& 1.13	\\
18.75 & 24.8   & 9.7   & 8.0	& 14	& 1.13	\\
19.25 & 34.9   & 11.7  & 10.1	& 20	& 1.13	\\
19.75 & 85   & 17  & 16		& 48	& 1.13	\\
20.25 & 137  & 24  & 23		& 78	& 1.13	\\
20.75 & 252  & 34  & 34		& 142	& 1.13	\\
21.25 & 360  & 51  & 49		& 138	& 0.79	\\
21.75 & 967  & 103 & 101	& 366	& 0.79	\\
22.25 & 1888 & 171 & 171	& 716	& 0.79	\\
22.75 & 3332 & 273 & 277	& 1238	& 0.79	\\
\cutinhead{FUV}
16.75 & 1.7	& 4.1 & 1.5	& 1	& 1.13  \\
17.25 & 3.6    & 4.8   & 2.4	& 2	& 1.13	\\
17.75 & 3.5    & 4.8   & 2.4	& 2	& 1.13	\\
18.25 & 8.8    & 6.3   & 4.2	& 5	& 1.13	\\
18.75 & 16.0   & 7.8   & 6.0	& 9	& 1.13	\\
19.25 & 10.3    & 6.6   & 4.6	& 6	& 1.13	\\
19.75 & 33.3   & 11.7   & 10.2	& 19	& 1.13	\\
20.25 & 83   & 19  & 18		& 47	& 1.13	\\
20.75 & 101  & 19  & 18		& 57	& 1.13	\\
21.25 & 168  & 34  & 32		& 65	& 0.79	\\
21.75 & 356  & 55  & 53		& 135	& 0.79	\\
22.25 & 710  & 87  & 86		& 271	& 0.79	\\
22.75 & 1344 & 140 & 140	& 503	& 0.79	\\
23.25 & 2153 & 202 & 202	& 793	& 0.79	\\
\enddata
\tablecomments{Magnitudes are reported for the center of the bins, differential number counts have units of mag$^{-1}$ deg$^{-2}$,
$N_{raw}$ are the raw galaxy counts in each bin (not including statistical corrections), and the coverage area is given for square degrees.}
\end{deluxetable}

\begin{deluxetable}{clccccccc}
\tablecolumns{9}
\tabletypesize{\scriptsize}
\tablecaption{\label{sdss_real} Confirmed Objects in the SDSS DR6 Catalog using HST-ACS}
\tablewidth{0pt}
\tablenum{3}
\tablehead{
\colhead{$r$} & \colhead{$N_{gal}$} & \colhead{$f$} & \colhead{$\Delta$$f_{u}$} & \colhead{$\Delta$$f_{l}$} & \colhead{$N_{star}$} & \colhead{$f$} & \colhead{$\Delta$$f_{u}$} & \colhead{$\Delta$$f_{l}$}\\
\colhead{(1)} & \colhead{(2)} & \colhead{(3)} & \colhead{(4)} & \colhead{(5)} & \colhead{(6)} & \colhead{(7)} & \colhead{(8)} & \colhead{(9)}}
\startdata
18.5 & 28  & 0.96 & 0.03 & 0.08 & 13 & 1.00 & 0.00 & 0.13 \\
19.5 & 56  & 0.95 & 0.03 & 0.05 & 23 & 1.00 & 0.00 & 0.08 \\
20.5 & 131 & 0.97 & 0.01 & 0.02 & 33 & 0.91 & 0.05 & 0.08 \\
21.5 & 247 & 0.86 & 0.02 & 0.03 & 33 & 0.91 & 0.05 & 0.08 \\
22.5 & 240 & 0.88 & 0.02 & 0.02 & 86 & 0.98 & 0.02 & 0.03 \\
23.5 & 66  & 0.89 & 0.04 & 0.05 & 72 & 0.78 & 0.05 & 0.06 \\
24.5 & 46  & 0.59 & 0.08 & 0.08 & 38 & 0.61 & 0.09 & 0.09 \\
\enddata
\end{deluxetable}
\begin{deluxetable}{clccclccc}
\tablecolumns{9}
\tabletypesize{\scriptsize}
\tablecaption{\label{sdss_real} Confirmed Galaxies Among SDSS Objects Classified as Galaxy or Star}
\tablewidth{0pt}
\tablenum{4}
\tablehead{
\colhead{$r$} & \colhead{$N_{gal}$} & \colhead{$g$} & \colhead{$\Delta$$g_{u}$} & \colhead{$\Delta$$g_{l}$} & \colhead{$N_{star}$} & \colhead{$g$} & \colhead{$\Delta$$g_{u}$} & \colhead{$\Delta$$g_{l}$} \\
\colhead{(1)} & \colhead{(2)} & \colhead{(3)} & \colhead{(4)} & \colhead{(5)} & \colhead{(6)} & \colhead{(7)} & \colhead{(8)} & \colhead{(9)}}	\\
\startdata
\cutinhead{ACS/SDSS Detections in the Coma-1 and Coma-3 Fields}
18.5 & 32  & 0.97 & 0.03 & 0.07 & 14  & 0.07 & 0.15 & 0.06 \\
19.5 & 78  & 0.95 & 0.02 & 0.04 & 33  & 0.03 & 0.07 & 0.03 \\
20.5 & 192 & 0.93 & 0.02 & 0.02 & 44  & 0.07 & 0.04 & 0.06 \\
21.5 & 299 & 0.94 & 0.01 & 0.02 & 62  & 0.32 & 0.07 & 0.06 \\
22.5 & 320 & 0.93 & 0.01 & 0.02 & 137 & 0.50 & 0.04 & 0.04 \\
23.5 & 86  & 0.90 & 0.03 & 0.04 & 83  & 0.66 & 0.06 & 0.06 \\
24.5 & 38  & 0.92 & 0.04 & 0.07 & 36  & 0.53 & 0.09 & 0.10 \\
\cutinhead{ACS/SDSS/GALEX NUV Detections in the Coma-3 Field}
18.5 & 6  & 0.83 & 0.14 & 0.29 & 1  & 0.00 & 0.84 & 0.00	\\
19.5 & 11 & 1.00 & 0.00 & 0.15 & 3  & 0.00 & 0.46 & 0.00	\\
20.5 & 31 & 1.00 & 0.00 & 0.06 & 4  & 0.00 & 0.37 & 0.00	\\
21.5 & 63 & 0.97 & 0.02 & 0.04 & 4  & 0.50 & 0.31 & 0.31	\\
22.5 & 61 & 0.98 & 0.01 & 0.04 & 13 & 0.77 & 0.12 & 0.18	\\
23.5 & 9  & 1.00 & 0.00 & 0.18 & 7  & 0.86 & 0.12 & 0.26	\\
\cutinhead{ACS/SDSS/GALEX FUV Detections in the Coma-3 Field}
18.5 & 4  & 0.75 & 0.21 & 0.37 & 0  & \nodata & \nodata & \nodata       \\
19.5 & 12 & 1.00 & 0.00 & 0.14 & 0  & \nodata & \nodata & \nodata        \\
20.5 & 25 & 1.00 & 0.00 & 0.07 & 1  & 0.00 & 0.00 & 0.84        \\
21.5 & 45 & 0.98 & 0.02 & 0.05 & 2  & 1.00 & 0.00 & 0.60        \\
22.5 & 27 & 0.96 & 0.03 & 0.08 & 5  & 1.00 & 0.00 & 0.31        \\
23.5 & 6  & 1.00 & 0.00 & 0.26 & 4  & 1.00 & 0.00 & 0.37        \\
\enddata
\end{deluxetable}

\end{document}